\documentclass[traditabstract]{aa}
\usepackage{natbib}
\bibpunct{(}{)}{;}{a}{}{,}
\usepackage{graphicx}
\usepackage{revsymb}    
\usepackage{amsmath}    
\usepackage[usenames]{color}    
\usepackage{epstopdf}   
\def\ind#1{_{\rm#1}}

\newcommand{\vt}{{\rm v}}

\newcommand{\derivp} [2] {\frac {\partial #1 } {\partial #2} }
\newcommand{\deriv} [2] {\frac {\textrm{d} #1 } {\textrm{d} #2} }
\newcommand{\derivD} [2] {\frac {\textrm{D} #1 } {\textrm{D} #2} }

\newcommand{\eq}[1] {Eq.\,(\ref{#1})}

\usepackage{txfonts}

\usepackage[switch]{lineno} 
\usepackage[normalem]{ulem}

\begin{document}

\title{Angular momentum redistribution by mixed modes
\\ in evolved low-mass stars}
\subtitle{I. Theoretical formalism}

\author{K. Belkacem\inst{1}, J.~P. Marques\inst{2}, M.~J. Goupil\inst{1}, T. Sonoi\inst{1}, R.~M. Ouazzani\inst{3}, M.~A. Dupret\inst{4}, S. Mathis\inst{5,1}, B. Mosser\inst{1}, \and M. Grosjean\inst{4}}

\institute{
 LESIA, Observatoire de Paris, PSL Research University, CNRS, Universit\'e Pierre et Marie Curie,
 Universit\'e Denis Diderot,  92195 Meudon, France
\and
Institut d'Astrophysique Spatiale, CNRS, Universit\'e Paris XI,
   91405 Orsay Cedex, France
\and
Stellar Astrophysics Centre, Department of Physics and Astronomy, Aarhus University, 
Ny Munkegade 120, DK-8000 Aarhus C, Denmark
\and
Institut d'Astrophysique et de G\'eophysique, Universit\'e de Li\`ege, All\'ee du 6 Ao\^ut 17-B 4000 Li\`ege, Belgium 
\and
Laboratoire AIM Paris-Saclay, CEA/DSM-CNRS-Universit\'e
Paris Diderot; IRFU /SAp, Centre de Saclay, 91191 Gif-sur-Yvette Cedex, France
}

   \offprints{K. Belkacem}
   \mail{kevin.belkacem@obspm.fr}
   \date{\today}

  \authorrunning{Belkacem}

   \abstract{
   Seismic observations by the space-borne mission \emph{Kepler} have shown  that the core of red giant stars slows down while evolving, requiring an efficient physical mechanism  to extract angular momentum from the inner layers. Current stellar evolution codes fail to reproduce the observed rotation rates by several orders of magnitude and instead predict a drastic spin-up of red giant cores. New efficient mechanisms of angular momentum transport are thus required.
   
   In this framework, our aim is to investigate the possibility that mixed modes extract angular momentum from the inner radiative regions of evolved low-mass stars. To this end, we consider the transformed Eulerian mean (TEM) formalism, which allows us to consider the combined effect of  both the wave momentum flux in the mean angular momentum equation and the wave heat flux in the mean entropy equation as well as their interplay with the meridional circulation. 
   In radiative layers of evolved low-mass stars, the quasi-adiabatic approximation, the limit of slow rotation, and the asymptotic regime can be applied for mixed modes and enable us to establish a prescription for the wave  fluxes in the mean equations. The formalism is finally applied to a $1.3 M_\odot$ benchmark model, representative of observed CoRoT and \emph{Kepler} oscillating evolved stars. 
    
    We show that the influence of the wave heat flux on the mean angular momentum is not negligible and that the overall effect of mixed modes is to  extract angular momentum from the innermost region of the star. A quantitative and accurate estimate requires realistic values of mode amplitudes. This is provided in a companion paper. }


   \maketitle

\section{Introduction}
\label{intro}

Rotation has important consequences on stellar evolution. In particular, it induces meridional circulations, as well as shear and baroclinic instabilities, which contribute to the redistribution of angular momentum and to the mixing of chemical elements \citep[see for instance][for comprehensive reviews]{Talon07,Maeder2009,Mathis2013b,Palacios2013}. Except for the Sun, the observational constraints  remained sparse until the advent of the space-borne missions CoRoT \citep{Baglin2006a,Baglin2006b,Michel2008} and \emph{Kepler} \citep{Borucki2010,Bedding2010,Chaplin2011}. 
 These were mainly constraints on the efficiency of the transport of chemicals \citep[e.g., ][]{Charbonnel2007} provided by observations of surface abundances. 

Based on seismic measurements, \cite{Beck2012} and \cite{Deheuvels2012,Deheuvels2014} brought stringent observational constraints on the rotation profiles in the innermost layers of subgiant stars observed by \emph{Kepler}. They concluded that the core of  subgiant stars spins up, while their envelope decelerates. Their core rotates about three to ten times faster than the envelope, depending on the evolutionary state of the star. On the other hand, \cite{Mosser2012} analysed a sample of about 300 red-giant stars observed by \emph{Kepler} and found that, surprisingly, the mean core rotation rate decreases significantly during the red-giant phase. 

Current models of red-giant stars including angular momentum redistribution processes are unable to explain such low core rotation rates in subgiant and  red giant stars. They are also unable to explain the deceleration of the core during the ascent of the red-giant branch. Indeed, \cite{Eggenberger2012}, \cite{Marques2013}, and \cite{Ceillier2013}  computed stellar models with transport of angular momentum by both meridional circulation and shear instabilities, concluding that these processes cannot explain the observed rotation profiles. The authors emphasised the need for an additional physical mechanism to do so. 
Along the same lines, \cite{Cantiello2014} included the rotationally induced circulations and shear instabilities together with the effect of a magnetic field generated through the Tayler-Spruit dynamo \citep{Spruit1999,Spruit2002} and reached the same conclusion. However, a recent work by \cite{Rudiger2015} has shown that magneto-rotational instabilities of a toroidal magnetic field could explain the angular momentum redistribution in subgiants and early red giants \citep[see also][]{Maeder2014}. 
Internal gravity waves\footnote{Note that all along this paper, we will distinguish between progressive waves (hereafter mentioned by \emph{waves}) and stationary waves (hereafter mentioned by \emph{modes}).} can also transport angular momentum \citep[e.g.,][]{Press1981} and it has been demonstrated that they could explain the nearly flat rotation profile in the inner radiative zone of the Sun \citep{Charbonnel2005}. While this process is still to be investigated for subgiant and red giant stars, \cite{Fuller2014} have found that internal gravity waves are likely to couple the convective region and the upper radiative region, but not the innermost layers. 
Hence, the question remains open, and we still need an efficient physical mechanism that explains the slow-down in the core of evolved low-mass stars. 

Our aim in this paper is to investigate the influence of normal modes (more precisely, mixed modes) on the core rotation of subgiant and red giant stars. Angular momentum can be transferred through energy exchanges between the oscillations and the mean flow, which results in a modification of the rotation profile. 
Non-radial modes in subgiants and red giants have two interesting properties. Firstly, they have large amplitudes both in the outer layers and in the core of the star so that they are efficiently excited by turbulent convection while having a non-negligible effect on the innermost layers. Secondly, their amplitudes can be inferred, since they have been observed by CoRoT and \emph{Kepler} \citep[e.g.,][]{Bedding2011,Mosser2011,Chaplin2013}. It is possible, therefore, to provide not only qualitative but also quantitative estimates of the angular momentum transport.   

The problem of the redistribution of angular momentum by progressive waves and normal modes has been considered for a long time from a theoretical point of view. For modes, the pioneer work of \cite{Ando1983} addressed the question of the interaction of wave and rotation through wave momentum stresses in the mean angular momentum equation. Following this work, efforts have been made to theoretically
address  the problem of the redistribution of angular momentum in classical pulsators by unstable modes \citep[e.g.,][]{Ando1986,Lee1993,Lee2007,Townsend2008,Townsend2014}. Nevertheless, a quantitative estimate is difficult to make since it remains a challenge to predict the amplitude of modes in classical pulsators \citep[see, however, ][]{Lee2014}.  
In parallel, \cite{Press1981,Zahn1997,Kumar1999,Mathis2008,Mathis2013} have considered the interaction of the wave with the mean flow for progressive internal gravity waves (IGW) with a slightly different formalism. Nevertheless, all these works mainly
focused on the effect of waves on the mean angular momentum equation and neglected their effects on the mean energy equation. 

In contrast, the interaction between waves/modes and meridional circulations has been extensively studied in geophysical flows in the 60s and 70s in the context of middle atmosphere dynamics \cite[e.g.,][]{Andrews1987,Holton1992}. \cite{Andrews76,Andrews78} proposed the \emph{\textup{transformed Eulerian mean}} formalism (TEM) to account for the interaction between waves and the mean flow in a general way. That allowed them to demonstrate the \emph{non-acceleration theorem}: if the waves are steady, conservative, and of small amplitudes, the mean flow is not accelerated. In the present series of papers, we adopt this formalism and adapt it to model angular momentum transport by mixed modes in stars. In this paper, which is the first of the series, we establish the formalism and discuss numerical results for a benchmark model. In a companion paper (hereafter paper II) we consider realistic mode amplitudes to quantitatively estimate the efficiency of the angular momentum transport by modes along the evolution of low-mass stars. 

This paper is  organised as follows: Section~\ref{wavemeanflow_eulerian} introduces the mean flow equations. Section~\ref{TEM} discusses the transformed Eulerian mean (TEM) formalism and its particular form in the case of stellar shellular rotation. In Sect.~\ref{sect_modes}, we model the wave fluxes for mixed modes. In Sect.~\ref{numericalresults}, the rate of angular momentum transport is computed numerically for a benchmark model of an evolved $1.3 M_\odot$ star, representative of CoRoT and \emph{Kepler} observations. Section~\ref{conclusions} is dedicated to discussions and conclusions. 

\section{Preliminary remarks on the wave mean-flow equations}
\label{wavemeanflow_eulerian}

In this section, we derive the equations that describe the interaction between the mean flow and waves. We use an Eulerian averaging process to derive the equations of the mean flow. It  allows us to discuss the coupling between waves and the meridional circulation as well as the role of the wave heat flux. 

The continuity, momentum, and energy equations in an inertial frame can be written as
\begin{align}
\label{continuity}
&\derivp{\rho}{t} + \vec \nabla \cdot \left( \rho \vec \vt \right) = 0 \\
\label{moment}
&\derivp{\left( \rho \vec \vt \right)}{t} + \vec \nabla \cdot \left( \rho \vec \vt \vec \vt \right) = - \vec \nabla p  - \rho \vec \nabla \Phi + \vec X \\
\label{energy}
&\derivp{\left( \rho s \right)}{t} + \vec \nabla \cdot \left( \rho \vec \vt s \right) = Q  \, , 
\end{align}
where $\rho$ is the density, $\vec \vt$ the velocity field, $p$ the pressure, $s$ the specific entropy, $\Phi$ the gravitational potential, $\vec X$ is a non-conservative mechanical forcing (e.g., turbulent dissipation), and $Q$ represents heating or cooling terms.  

The azimuthal component of \eq{moment} can be written in a form that represents the conservation of specific angular momentum:
\begin{align}
\label{momentum}
\derivp{\left( \rho h \right)}{t} + \vec \nabla \cdot \left(  \rho h \vec \vt \right) = - \derivp{p}{\phi} - \rho \derivp{\Phi}{\phi} + \varpi \, X_\phi \, , 
\end{align}
where the specific angular momentum is $h=\varpi \, \vt_\phi = \varpi^2 \, \Omega$, with $\Omega$  representing the angular velocity, $\varpi = r \sin \theta$, and $\vt_\phi$  the azimuthal component of the velocity field. 

A given field can be decomposed into a mean part and a perturbation. The perturbation is associated with linear non-radial waves. More precisely, for a given field $A$, we have
\begin{equation}
\label{decomposition_scalar}
A = \overline{A} + A^\prime,
\end{equation}
where $\overline{A}$ is the Eulerian-mean azimuthal average 
\begin{equation}
 \overline{A} = \frac{1}{2 \pi} \int_0^{2\pi} A \; {\rm d}\phi \, .
\end{equation}

Introducing the decomposition given by \eq{decomposition_scalar} into Eqs.~(\ref{continuity}), (\ref{energy}), and (\ref{momentum}) and azimuthal averaging them, we obtain the following system:
\begin{align}
 \label{continuity_eulerian_tmp}
 &\derivp{\overline{\rho}}{t}  
 + \vec \nabla_\bot \cdot \left( \overline{\rho} \, \overline{\vec \vt}_\bot \right)
 = \mathcal{D} \\
\label{momentum_eulerian_tmp}
&\overline{\rho} \derivp{\overline{h}}{t}   
+ \overline{\rho} \, \left( \overline{\vec \vt}_\bot \cdot \vec \nabla_\bot \right) \overline{h} 
= - \vec \nabla_\bot \cdot \left( \varpi \overline{\rho} \, \overline{ \vt_\phi^\prime \vec \vt^\prime_\bot} \right) 
+ \varpi  \overline{X}_\phi + \mathcal{H} \\
\label{energy_eulerian_tmp}
&\overline{\rho} \derivp{\overline{s}}{t} + 
\overline{\rho}\left( \overline{\vec \vt}_\bot \cdot \vec \nabla_\bot \right) \overline{s} 
= - \vec \nabla_\bot \cdot \left(\overline{\rho} \, \overline{s^\prime \vec \vt^\prime_\bot} \right) + \overline{Q} + \mathcal{S,}
\end{align}
with
\begin{align}
\label{term_h}
\mathcal{H} &=  - \overline{\rho^\prime \vec \vt^\prime_\bot} \cdot \vec \nabla_\bot  \overline{h} - \varpi \derivp{\overline{\rho^\prime \vt_\phi^\prime}}{t} - \overline{\rho^\prime \derivp{\Phi^\prime}{\phi}} 
 - \vec \nabla_\bot \cdot \left( \varpi \overline{\rho^\prime \vt_\phi^\prime} \, \overline{\vec \vt}_\bot \right), \\ 
\label{term_s}
\mathcal{S} &= -  \overline{\rho^\prime \vec \vt^\prime_\bot} \cdot \vec \nabla_\bot \overline{s}
- \vec \nabla_\bot \cdot \left(\overline{\rho^\prime s^\prime} \, \overline{\vec \vt}_\bot \right) - \derivp{\overline{ \rho^\prime s^\prime}}{t} ,\\
\label{term_d}
\mathcal{D} &= - \vec \nabla_\bot \cdot \left(\overline{\rho^\prime \, \vec \vt^\prime_\bot} \right) ,
\end{align}
and $\vec \nabla_\bot$, $\vec \nabla_\bot \cdot \left(\right)$, $\vec \vt_\bot$ are the gradient, divergence, and velocity vector in the meridional plane, respectively. They are defined by
\begin{align}
\vec \nabla_\bot = \vec e_r  \derivp{}{r}  + \vec e_\theta\frac{1}{r} \derivp{}{\theta}   ,
\quad {\rm and} \quad \vec \vt_\bot = \vt_r \, \vec e_r + \vt_\theta \, \vec e_\theta \, .
\end{align}

The terms involving the density perturbation ($\rho^\prime$) are generally considered to be small \citep[e.g.,][]{Ando1983,Unno89}. This is particularly the case for low-frequency waves ($\sigma_R \ll N$, where $\sigma_R$ is the wave frequency and $N$ the buoyancy frequency) where the anelastic approximation applies \citep[e.g.,][]{Dintrans2001}. Therefore, we neglect the terms $\overline{\rho^\prime \, \vec \vt^\prime_\bot}$, $\overline{\rho^\prime s^\prime}$, and $\overline{\rho^\prime \, \vt_\phi^\prime}$ in Eqs.~(\ref{term_h}) to (\ref{term_d}). In addition, we use the Cowling approximation by neglecting the perturbation of the gravitational potential, so that Eqs.~(\ref{continuity_eulerian_tmp}) to (\ref{energy_eulerian_tmp}) become
\begin{align}
 \label{continuity_eulerian}
& \derivp{\overline{\rho}}{t}  
 + \vec \nabla_\bot \cdot \left( \overline{\rho} \, \overline{\vec \vt}_\bot \right)
 = 0 \\
\label{momentum_eulerian}
&\overline{\rho} \derivp{\overline{h}}{t}   
+ \overline{\rho} \, \left( \overline{\vec \vt}_\bot \cdot \vec \nabla_\bot \right) \overline{h} 
= - \vec \nabla_\bot \cdot \left( \varpi \overline{\rho} \, \overline{ \vt_\phi^\prime \vec \vt^\prime_\bot} \right)
 + \varpi \overline{X}_\phi   \\
\label{energy_eulerian}
&\overline{\rho} \derivp{\overline{s}}{t} + 
\overline{\rho}\left( \overline{\vec \vt}_\bot \cdot \vec \nabla_\bot \right) \overline{s} 
= - \vec \nabla_\bot \cdot \left(\overline{\rho} \, \overline{s^\prime \vec \vt^\prime_\bot} \right) 
 + \overline{Q}  \, , 
\end{align} 
where the effect of waves now appears through the wave momentum flux in the mean angular momentum equation, Eq. \eqref{momentum_eulerian} and the wave heat flux in the mean entropy equation, Eq. \eqref{energy_eulerian}. 

However, the mean specific entropy $\overline{s}$ and mean specific angular momentum $\overline{h}$ are not independent. They are connected by the baroclinic equation (also known as the thermal wind balance equation), obtained by taking the curl of the hydrostatic equilibrium equation:
\begin{align}
\label{baroclinic_eulerian}
&\overline{\rho}^2 \; \vec \nabla_\bot \left(\frac{\overline{h}}{\varpi^2} \vt_\phi  \right) \times \vec \nabla_\bot \varpi + \vec \nabla_\bot \overline{\rho}  \times \vec \nabla_\bot \overline{p} = 0 \, , 
\end{align}
together with the equation of state $\overline{s} = \overline{s} \, (\overline{\rho},\overline{p})$. 

Therefore, we must solve the complete set of Eqs.~\eqref{continuity_eulerian} to \eqref{baroclinic_eulerian} to compute the effect of waves on the mean flow, including both wave stresses in the mean angular momentum equation,~ Eq. \eqref{momentum_eulerian}, and the wave heat flux in the mean entropy equation, Eq.~\eqref{energy_eulerian}. However, the effect of the wave heat flux is often neglected in the literature related to transport of angular momentum in stars by waves \citep[e.g.,][]{Press1981,Ando1983,Lee1993,Zahn1997,Kumar1999,Pantillon2007,Mathis2013,Townsend2014}.

It is potentially misleading to overlook the effect of the wave heat flux. To illustrate this point, we consider the extreme case of a conservative and stationary wave field. Under these assumptions, the \emph{\textup{non-acceleration theorem}} states that the mean flow is not accelerated by waves, although the wave momentum and heat fluxes defined in Eqs.~\eqref{momentum_eulerian} and \eqref{energy_eulerian} do not vanish. 
In other words, the wave fluxes of heat and momentum produce a meridional circulation that cancels their tendency to affect the mean flow. Indeed, the wave momentum and wave heat fluxes are not independent and, in the presence of wave forcing, the meridional circulation must ensure that $\overline{h}$ and $\overline{s}$ satisfy the baroclinic equation. 

Consequently, both the wave momentum and wave heat fluxes must be considered to account for the effect of waves on the mean flow. To do so, we  adopt in the following section the TEM formalism introduced by \cite{Andrews76,Andrews78}. 
 
\section{Transformed Eulerian mean equations}
\label{TEM}

\subsection{Formalism}

The wave heat flux $\vec R = \overline{s^\prime \vec \vt^\prime_\bot}$ can be split into a component along an isentropic surface, the
skew flux, and a component perpendicular to it. Following \cite{Vallis2006},
\begin{align}
\vec R = \left( \vec n \times \vec R \right) \times \vec n + \left(\vec n \cdot \vec R \right) \, \vec n \;, 
\end{align}
with $\vec n = \vec \nabla_\bot \overline{s} / \vert \vec \nabla_\bot \overline{s} \vert$.
The divergence of the skew flux can be rewritten as
\begin{align}
\label{split_div_flux}
\vec \nabla_\bot \cdot \left[ \left( \vec n \times \vec R \right) \times \vec n \right] &= 
\vec \nabla_\bot \cdot \left( \frac{  \vec \nabla_\bot \overline{s} \times \vec R }{\vert \vec \nabla_\bot \overline{s} \vert^2} \times \vec \nabla_\bot \overline{s} \right) \nonumber \\
&= \left( \vec \nabla_\bot \times \frac{\vec \nabla_\bot \overline{s} \times \vec R}{\vert \vec \nabla_\bot \overline{s} \vert^2} \right) \cdot \vec \nabla_\bot \overline{s} \nonumber \\
&= \tilde{\vec \vt}\cdot \vec \nabla_\bot \overline{s}\, ,
\end{align}
where we used the relation $\vec \nabla \cdot (\vec a \times \vec \nabla \alpha) = \vec \nabla \alpha \cdot (\vec \nabla \times \vec a)$, valid for any scalar $\alpha$ and vector $\vec a$. 

Equation~(\ref{split_div_flux}) shows that the skew flux behaves like an advection by the velocity $\tilde{\vec \vt}$. The main motivation underlying the TEM is thus to incorporate the advective part of the wave heat flux into the mean meridional velocity field ($\overline{\vec \vt}$). 

The \emph{residual meridional circulation} is thus defined by  
\begin{align}
\label{residual_v}
 \overline{\rho} \, \overline{\vec \vt}_{\bot}^\dag =  \overline{\rho} \, \overline{\vec \vt}_\bot  + \vec \nabla_\bot \times \left(\overline{\rho} \, \overline{\psi} \, \vec e_\phi\right) \, , 
\end{align}
where the stream function $\overline{\psi}$ is defined, from \eq{split_div_flux}, by 
 \begin{align}
 \label{def_psi_general}
\overline{\psi} =  \frac{\vec \nabla_\bot \overline{s} \times \vec R}{\vert \vec \nabla_\bot \overline{s} \vert^2} \cdot \vec e_\phi= \frac{1}{\vert \vec \nabla_\bot \overline{s} \vert^2} \left[ \left(\derivp{\overline{s}}{r}\right) \, \overline{\vt_\theta^\prime s^\prime} - \frac{1}{r} \left(\derivp{\overline{s}}{\theta}\right) \, \overline{\vt_r^\prime s^\prime} \right]  \, .
\end{align}
We note that the definition for $\psi$ given in \eq{def_psi_general} is not unique \citep[see][for details]{Plumb2005,Vallis2006}. 

Inserting Eqs.~(\ref{residual_v}) and (\ref{def_psi_general}) into \eq{continuity_eulerian} to \eq{energy_eulerian}, we have 
\begin{align}
 \label{continuity_eulerian_final}
 \derivp{\overline{\rho}}{t}  
 + \vec \nabla_\bot \cdot \left( \overline{\rho} \, \overline{\vec \vt}_\bot^\dag \right)
 &= 0 \\
\label{momentum_eulerian_final}
\overline{\rho} \derivp{\overline{h}}{t}   
+ \overline{\rho} \, \left( \overline{\vec \vt}_\bot^\dag \cdot \vec \nabla_\bot \right) \overline{h} 
&= - \vec \nabla_\bot \cdot \left( \overline{\rho} \, \vec F \right) + \varpi \overline{X}_\phi  \\
\label{energy_eulerian_final}
\overline{\rho} \derivp{\overline{s}}{t} + 
\overline{\rho}\left( \overline{\vec \vt}_\bot^\dag \cdot \vec \nabla_\bot \right) \overline{s} 
&= - \vec \nabla_\bot \cdot \left( \overline{\rho}\, \vec G \right)
 + \overline{Q}  \, , 
\end{align}
where the components of the vectors $\vec F$ and $\vec G$ are given by
\begin{align}
\label{waves_fluxes}
 F_r &= \varpi \, \overline{\vt_\phi^\prime \, \vt_r^\prime} + \frac{\overline{\psi} }{r} \, \derivp{\overline{h}}{\theta} \; , \quad 
 F_\theta = \varpi \, \overline{\vt_\phi^\prime \, \vt_\theta^\prime} - \overline{\psi} \, \derivp{\overline{h}}{r} \, , \\
 G_r &=  \overline{s^\prime \, \vt_r^\prime} + \frac{\overline{\psi} }{r} \, \derivp{\overline{s}}{\theta}  \; , \quad
 G_\theta =  \overline{s^\prime  \, \vt_\theta^\prime} - \overline{\psi} \, \derivp{\overline{s}}{r} \, .
\end{align}
Equation~\eqref{baroclinic_eulerian} is left unmodified. 

Finally, when $\vec F$ and $\vec G$ are specified, the residual velocity $\overline{\vec \vt}_\bot^\dag$ becomes part of the solution of Eqs.~\eqref{continuity_eulerian_final} to \eqref{energy_eulerian_final} together with Eq.~\eqref{baroclinic_eulerian}. They are  strictly equivalent to Eqs.~\eqref{continuity_eulerian} - \eqref{energy_eulerian}, but present several conceptual and practical advantages:

\begin{itemize}
\item The TEM equations enable distinguishing the advective and diffusive parts of the wave heat flux and incorporating the advective component in the mean velocity field. 
\item  \cite{Andrews78} showed that $\vec \nabla \cdot  \left(\overline{\rho} \vec F\right)$ and $\vec \nabla \cdot \left(\overline{\rho} \vec G\right)$  only depend on wave dissipation and non-steady terms. It is the \emph{non-acceleration theorem}. Therefore, the TEM makes the adiabatic and non-adiabatic contributions of waves more explicit, the latter being able to modify the mean flow. 
\item Finally, depending upon the considered problem, the TEM formalism allows us to gather in a single equation and in the single term $\vec F$ both the wave momentum and wave heat fluxes. It explicitly shows that wave momentum fluxes and wave heat fluxes do not influence the mean flow separately, but only in the combination given by $\vec F$. For instance, in the quasi-geostrophic approximation  $\vec G \approx 0$ so that the term $ \vec F$ (the Eliassen-Palm flux) is the only wave contribution to the problem. As we will show in the companion paper,  we are in the same situation for mixed modes in the limit of shellular rotation, in which the radial component of $\vec G$ can be neglected. 
\end{itemize}

\subsection{Equation for the mean angular momentum: shellular rotation}
\label{shellular}

Following the work of 
\cite{Zahn1992,MaederZahn1998,Mathis2004} \citep[see also][for an extensive discussion]{Maeder2009}, shellular rotation has been an approximation widely used in 1D stellar evolutionary codes. It is based on the assumption that turbulence is highly anisotropic in stellar radiative zones \citep[e.g.,][]{Talon1997,Maeder2003,Mathis2004b}; it is much stronger in the horizontal than in the vertical direction. Accordingly, efficient horizontal turbulent viscosity ensures that the angular velocity is almost constant on isobars. We then consider the TEM equations in this framework. 

Given that the specific angular momentum is defined by $\overline{h} = \varpi^2\, \overline{\Omega} (r,\theta)$, we introduce the following decomposition
\begin{align}
\label{shellular_omega}
\overline{\Omega} (r,\theta)= \Omega_0 \left(r \right) + \widehat{\Omega} \left(r,\theta\right)
\, , 
\end{align}
with
\begin{align}
\Omega_0   = \frac{ \int_{0}^{\pi} \sin^3 \theta \; \overline{\Omega}(r,\theta)\; {\rm d} \theta}{ \int_{0}^{\pi} \sin^3 \theta \; {\rm d} \theta} = \frac{3}{4} \int_{0}^{\pi} \sin^3 \theta \; \overline{\Omega}(r,\theta)\; {\rm d} \theta ,
\end{align}
where, with the shellular approximation, $\Omega_0 \gg \widehat{\Omega}$. 
The scalar quantities are developed as 
\begin{align}
\label{def_scalar} 
\overline{X}(r,\theta) &=  \left< X \right> \left(r \right) + \widehat{X} \left(r,\theta\right) 
\, , 
\end{align} 
with
\begin{align}
\left< X \right>   = \frac{ \int_{0}^{\pi} \sin \theta \; \overline{X}(r,\theta) \, {\rm d} \theta}{ \int_{0}^{\pi} \sin \theta \, {\rm d} \theta} = \frac{1}{2} \int_{0}^{\pi} \sin \theta \; \overline{X}(r,\theta) \, {\rm d} \theta ,
\end{align}
and the velocity field equals to $\overline{\vec \vt}_\bot^\dag \left(r,\theta\right) =  \dot{r} \, \vec e_r + \vec U^\dag \left(r,\theta\right)$, where $\dot{r}$ comes from the contraction or dilatation of the star and $\vec U^\dag$ is the residual meridional circulation. 

The various physical quantities are expanded in Legendre polynomials 
\begin{align}
\label{def_vitesse} 
\vec U^\dag \left(r,\theta\right) &= \sum_\ell \left[ U_\ell^\dag (r) \, P_\ell (\cos \theta) \, \vec e_r+ V^\dag_\ell (r) \deriv{P_\ell (\cos \theta)}{\theta} \, \vec e_\theta \right] \\
\widehat{\Omega} \left(r,\theta\right) &= \sum_\ell \Omega_\ell Q_\ell \left(\cos \theta\right) \\
\label{def_scalar2} 
\widehat{X} \left(r,\theta\right)  &= \sum_\ell  X_\ell (r) P_\ell \left(\cos \theta\right) \, , 
\end{align} 
where $Q_\ell = P_\ell - \delta_{\ell,0} +  \delta_{\ell,2} / 5$ \citep[see][]{Mathis2004}, and, for slow to moderate rotation, we assume that $\widehat{X} \ll  \left< X \right>$.

Following these assumptions, the specific entropy is nearly constant on isobars 
and the stream function $\overline{\psi}$ (Eq.~\ref{def_psi_general}) becomes at dominant order
 \begin{align}
 \label{def_psi_shellular}
\overline{\psi} = \left(\deriv{\left< s\right>}{r}\right)^{-1}  \, \overline{\vt_\theta^\prime s^\prime} \, ,
\end{align}
which is the same expression as for geostrophic flows \citep[e.g.,][]{Holton1992}. 

Introducing Eqs.~(\ref{def_vitesse}) -- (\ref{def_scalar2}) into \eq{momentum_eulerian_final}, horizontal averaging, neglecting the time-variation of the density, and using the expression of the residual stream function given by \eq{def_psi_shellular}, we obtain, at dominant order,
\begin{align}
\label{shellular_omega_r}
&\left< \rho \right> \deriv{\left(r^2 \Omega_0\right)}{t}   
= -\frac{1}{r^2} \derivp{}{r} \left[ r^2 \left( \mathcal{F}_{\rm circ} +  \mathcal{F}_{\rm shear} + \mathcal{F}_{\rm waves} \right) \right]
\end{align}
with the fluxes defined by
\begin{align}
\label{shellular_omega_r_advection}
\mathcal{F}_{\rm circ} &=  -\frac{1}{5} \left< \rho \right> r^2 \Omega_0 \, U_2^\dag \\
\label{shellular_omega_r_shear}
\mathcal{F}_{\rm shear} &= -\left< \rho \right> \nu\ind{v} r^2 \derivp{\Omega_0}{r} \\
\label{shellular_omega_r_wave}
\mathcal{F}_{\rm waves} &= \left< \rho \right>  \left< \varpi \left[ \overline{\vt_\phi^\prime \, \vt_r^\prime} + 2 \cos \theta \,  \Omega_0 \, \overline{\vt_\theta^\prime s^\prime} \left(\deriv{\left< s\right>}{r}\right)^{-1} \right] \right> \, .
\end{align}
The turbulent stresses are described by an anisotropic eddy viscosity denoted $\nu\ind{v}$ and $\nu\ind{h}$ for the vertical and horizontal component, respectively \citep[see][for details]{Zahn1992,Mathis2004}. 
We also note that we introduced the Lagrangian derivative ${\rm d} / {\rm d}t =  \partial / \partial t +  \dot{r} \, \partial / \partial r$.

Under the assumptions used for \eq{shellular_omega_r}, we obtain for the entropy equation  
\begin{align}
\label{energy_shellular_vertical}
 \left< \rho \right> \, \deriv{ \left< s \right>}{t}  = -\frac{1}{r^2} \derivp{}{r} \left< r^2 \,\mathcal{S}\right> + \left< Q \right> \, 
\end{align}
with 
\begin{align}
\label{energy_shellular_vertical_flux}
\mathcal{S} = \left< \rho\right>\,  \overline{s^\prime \, \vt_r^\prime} \, , 
\end{align}
and
\begin{align}
 \left< T\right> \left< Q \right> = \left< \rho \varepsilon \right> + \frac{ 1}{r^2} \frac{\partial}{\partial r}\left(r^2 \left< \chi \right> \frac{\partial \left<T\right>}{\partial r}\right)\, ,
\end{align}
where $T$ is the temperature, $\varepsilon$ is the nuclear energy generation rate, and $\chi$ is the thermal  conductivity (see, e.g., Sect. 6 of \citealt{Mathis2004}, and also \citealt{MaederZahn1998}). This expresses energy conservation on a level surface. To obtain \eq{energy_shellular_vertical}, we used \eq{def_psi_shellular} and assumed that the horizontal derivative of entropy is negligible at dominant  order. 

\subsection{Discussion}

As  mentioned in Sect.~\ref{wavemeanflow_eulerian}, transport of angular momentum by waves in stars has been extensively studied, but often ignoring the effect of the wave heat flux \citep[e.g.,][]{Press1981,Ando1983,Lee1993,Zahn1997,Kumar1999,Townsend2014}. Following the work of \cite{Bretherton1969}, \cite{Pantillon2007}  have also considered an additional wave contribution in the momentum equation, equal to $2 \cos \theta \, \Omega_0\, \overline{\vt_r^\prime \xi_\theta}$ in our notation \citep[see also][]{Lee2008,Mathis2009}. This term corresponds to a Lagrangian correction that can be understood as follows: on the typical time-scale of the momentum transport by waves (the modal period, for example), rotation advects a wave-related momentum flux. This term arises naturally in Lagrangian mean wave-flow theories such as the \emph{\textup{generalized Lagrangian mean}} (GLM) formulation by \cite{Andrews1978b}  \cite[see also][]{Grimshaw1984,Lee2013}.

For linear waves, one can show that this Lagrangian term can be recovered from the TEM equations if one assumes that the Lagrangian perturbation of entropy vanishes (\emph{i.e.} $\delta s =0$). The wave flux in the mean angular momentum equation (Eq.~\ref{shellular_omega_r}) is 
\begin{align}
\label{source_discussion}
\overline{\vt_\phi^\prime \, \vt_r^\prime} + 2 \cos \theta \,  \Omega_0 \, \overline{\vt_\theta^\prime s^\prime} \left(\deriv{\left< s\right>}{r}\right)^{-1} \, .
\end{align}

We can then express the entropy Eulerian perturbation in terms of the Lagrangian perturbation at first order: $\delta s = s^\prime + \xi_r \, {\rm d} \left< s \right> / {\rm d}r$, where $\xi_r$ is the wave displacement. Using \eq{def_perturbation_main} and the expression for the wave velocity $\vec \vt^\prime = i \left( \sigma_R + m\Omega_0 \right) \vec \xi - \varpi \left( \vec \xi \cdot \vec \nabla \Omega_0 \right) \, \vec e_\phi$ (where $\sigma_R$ is the modal frequency), we obtain
\begin{align}
\overline{\vt_\theta^\prime \, \xi_r} = - \overline{\vt_r^\prime \, \xi_\theta} \, .
\end{align}
Therefore, if the entropy Lagrangian perturbation vanishes (\emph{i.e.} $\delta s = 0$), the term (\ref{source_discussion}) becomes 
\begin{align}
\label{correction_breterthon69}
 \overline{\vt_\phi^\prime \, \vt_r^\prime} + 2 \cos \theta \,  \Omega_0 \, \overline{\vt_r^\prime \xi_\theta}  \, , 
 \end{align}
which is the expression given in \cite{Pantillon2007} as claimed. Finally, \eq{correction_breterthon69} shows that the Lagrangian correction introduced by \cite{Pantillon2007,Lee2008,Mathis2009} is recovered from the TEM equations. In this framework, it can be understood as originating from the wave heat source term in the mean entropy equation. 

However, the contribution related to the entropy Lagrangian perturbation is not negligible (see the results presented in Sect.~\ref{results_one_mode} below). Therefore, it is not enough to consider the wave flux in the mean angular momentum given by \eq{correction_breterthon69}.

\section{Modelling the wave fluxes: \emph{\textup{the case of mixed modes in evolved low-mass stars}}}
\label{sect_modes}

In this section, we focus on the transport of angular momentum by mixed normal modes  in the radiative region of evolved low-mass stars (i.e. subgiants and red giants). These modes have a dual nature; they behave as acoustic modes in the upper layers and as gravity modes in the inner layers. Therefore, they are detectable at the stellar surface while providing information on the innermost regions. They have been detected in several thousands of evolved stars \citep[e.g.,][]{Bedding2011,Mosser2012,Mosser2014} and have provided a wealth of information on the core of red giants \cite[e.g.,][]{Mosser2012}. These modes must not be confused with progressive gravity waves, often called \emph{\textup{internal}} gravity waves in the literature.  

\subsection{Simplifying approximations}
\label{wave_approximations}

We restrict ourselves to the case of strictly shellular rotation (see Sect.~\ref{shellular}). Moreover, we focus on the inner radiative regions since our aim in this series of papers is to investigate the transport of angular momentum in the core of evolved low-mass stars. These restrictions allow us to use several approximations  to describe the wave field: 
\begin{enumerate}
\item \emph{The quasi-adiabatic approach:} It consists of neglecting the difference between adiabatic and non-adiabatic eigenfunctions in the full wave equations. 
This approximation is valid when the local thermal time-scale is much longer than the modal period. This is the case in the radiative region of evolved low-mass  stars. 
\item \emph{The low-rotation limit:} We assume that the modal period is much shorter than the rotation period. This is justified by recent inferences of the rotation in the core of subgiants \citep{Deheuvels2012,Deheuvels2014} and red giants \citep{Mosser2012} using seismic constraints from \emph{Kepler}. 
\item \emph{The asymptotic limit:} Finally, we use an asymptotic description for gravity modes \citep[e.g.,][]{Dziembowski2001,Godart2009}, which is valid for mixed modes in the inner radiative region of subgiants and red giants \citep[e.g.,][]{Goupil2013}. 
\end{enumerate}

\subsection{Derivation of the wave fluxes}
\label{sourceterms}

We assume that the perturbations consist of non-radial oscillations such that, for any perturbed variable $f^\prime$, we have
\begin{align}
\label{def_perturbation_main}
f^\prime = a \, \mathcal{R}_e \left\{ \tilde{f}^\prime(r,\theta) \; e^{i (\sigma_R t + m \phi)}\right\},
\end{align} 
where  $\mathcal{R}_e$ denotes the real part, $a$ is the amplitude, and $\sigma_R$ is the real frequency (so that non-adiabatic effects are neglected). The amplitude is considered as statistically constant in time since we deal with solar-like oscillations resulting from a balance between mode driving and damping \citep[see][for details]{Samadi11}.   

We now consider the equations governing non-radial and non-adiabatic oscillations in the presence of rotation \citep[e.g., Eqs. 36.12 to 36.15 of][]{Unno89}. For sake of simplicity, tildes, bars, and brackets will be dropped from now on unless necessary. The equations are
\begin{align}
\label{eq1}
&\derivD{\vec \vt_\bot^\prime}{t} - \frac{2 h}{\varpi^2} \vt_\phi^\prime \nabla_\bot \varpi = -\frac{1}{\rho} \nabla_\bot p^\prime + \frac{\rho^\prime}{\rho^2} \nabla_\bot p \\
\label{eq2}
& \derivD{\vt_\phi^\prime}{t} +\frac{1}{\varpi} \vec \vt_\bot^\prime\cdot \nabla_\bot h = -\frac{i m}{\varpi} \frac{p^\prime}{\rho} \\
\label{eq3}
&\derivD{}{t} \left(\frac{\rho^\prime}{\rho}-\frac{p^\prime}{\Gamma_1 p} +\rho_T \frac{\delta s}{c_p}  \right) = - \vec \vt_\bot^\prime \cdot \vec A \\
\label{eq4}
&\derivD{\rho^\prime}{t} + \nabla_\bot \cdot \left(\rho \vec \vt_\bot^\prime \right) + \frac{i m \rho}{\varpi} \vt_\phi^\prime = 0\, ,
\end{align}
where $\Gamma_1=\left(\partial \ln p / \partial \ln \rho \right)_s$, $\rho_T=-\left(\partial \ln \rho / \partial \ln T \right)_p$, and 
\begin{align}
\derivD{}{t} = \derivp{}{t} + \Omega_0 \, \derivp{}{\phi} \, , \quad
\vec A = \vec \nabla \ln \rho - \frac{1}{\Gamma_1} \vec \nabla \ln p \, , 
\end{align}
so that $N^2=\rho^{-1} \vec \nabla p \cdot \vec A$. 

Given the assumption that the mode amplitude is statistically constant in time, Eqs.~(\ref{eq1}) to (\ref{eq4}) reduce to
\begin{align}
\label{eq1_stationaire}
&i (\sigma_R + m \Omega_0) \, \vec \vt_\bot^\prime - \frac{2 h}{\varpi^2} \vt_\phi^\prime \nabla_\bot \varpi = -\frac{1}{\rho} \nabla_\bot p^\prime + \frac{\rho^\prime}{\rho^2} \nabla_\bot p \\
\label{eq2_stationaire}
& i (\sigma_R + m \Omega_0) \, \vt_\phi^\prime +\frac{1}{\varpi} \vec \vt_\bot^\prime \cdot \nabla_\bot h = -\frac{i m}{\varpi} \frac{p^\prime}{\rho} \\
\label{eq3_stationaire}
&i (\sigma_R + m \Omega_0) \, \left(\frac{\rho^\prime}{\rho}-\frac{p^\prime}{\Gamma_1 p} +  \rho_T \frac{\delta s}{c_p}  \right) = - \vec \vt_\bot^\prime \cdot \vec A \\
\label{eq4_stationaire}
&i (\sigma_R + m \Omega_0) \, \rho^\prime + \nabla_\bot \cdot \left(\rho \vec \vt_\bot^\prime \right) + \frac{i m \rho}{\varpi} \vt_\phi^\prime = 0 \, . 
\end{align}

Using Eqs.~(\ref{eq1_stationaire}) to (\ref{eq4_stationaire}) and the assumptions detailed in Sect.~\ref{wave_approximations}, and after stirring vigorously as detailed in Appendix~\ref{AppA}, we obtain for a mode of a given angular degree $\ell$ and azimuthal order $m$ an expression of the form  
\begin{align}
\label{final_F_waves}
&-\frac{1}{r^2} \derivp{}{r} \left( r^2 \mathcal{F}_{\rm waves} \right) = \\
&a_{\ell,m}^{2} \left[ \mathcal{A}_{\ell}^m \; \derivp{^2\left(r^2 \Omega_0 \right)}{r^2} +\mathcal{B}_{\ell}^m \; \derivp{\left(r^2 \Omega_0 \right)}{r} + \mathcal{C}_{\ell}^m \; r^2 \Omega_0 + m \hat \sigma \, \mathcal{D}_{\ell}^m \right] \nonumber \, , 
\end{align}
where $\hat \sigma = \sigma_R + m\Omega_0$ and the coefficients $\mathcal{A}_{\ell}^m, \mathcal{B}_{\ell}^m, \mathcal{C}_{\ell}^m, \mathcal{D}_{\ell}^m$ are given by Eqs.~(\ref{coeffAf}) to (\ref{coeffDf}), respectively. The amplitude $a_{\ell,m}$ corresponds to the amplitude of a mode of angular degree $\ell$, and azimuthal order $m$. 

The derivation of the wave fluxes in \eq{energy_shellular_vertical} is more direct. Using the asymptotic expression for the Lagrangian perturbation of entropy given by \eq{deltaS_final}, we obtain 
\begin{align}
\label{final_T_2}
\frac{1}{r^2} \derivp{}{r} \left< r^2 \,\mathcal{S}\right> = \frac{a_{\ell,m}^{2}}{2 r^2} \derivp{}{r} \left( r^2 \, \rho \, \alpha \deriv{s}{r} k_r^2 \, \vert \xi_r^{\ell,m}\vert^2  \right) \, ,
\end{align}
where 
\begin{align}
\label{alpha_main}
&\alpha = -\frac{L}{4 \pi r^2 \rho T} \left( \frac{\nabla_{\rm ad}}{\nabla}-1 \right) \, \left(\deriv{s}{r}\right)^{-1}\\
\label{kr2_main}
&k_r^2 \simeq \frac{\ell(\ell+1)}{r^2} \, \left(\frac{N^2}{\sigma_R^2}-1\right) \, , 
\end{align}
with $L$ being the luminosity, $N$ the buoyancy frequency, $\nabla$ and $\nabla_{\rm ad}$  the actual and adiabatic temperature gradients, respectively. 

\section{Case of an evolved $1.3$ M$_\odot$ star}
\label{numericalresults}

The aim of this section is to qualitatively discuss the effect of mixed modes on the mean angular momentum profile within the TEM framework. To this end, we consider a stellar model for an evolved $1.3 M_\odot$ star that is representative of the oscillating red giant stars observed by CoRoT and \emph{Kepler}. A quantitative estimate as well as a discussion of the effect of mixed modes throughout the evolution is given in paper II. 

\subsection{Computing the benchmark model}
\label{models}

\begin{figure}[t]
\begin{center}
\includegraphics[width=9.2cm]{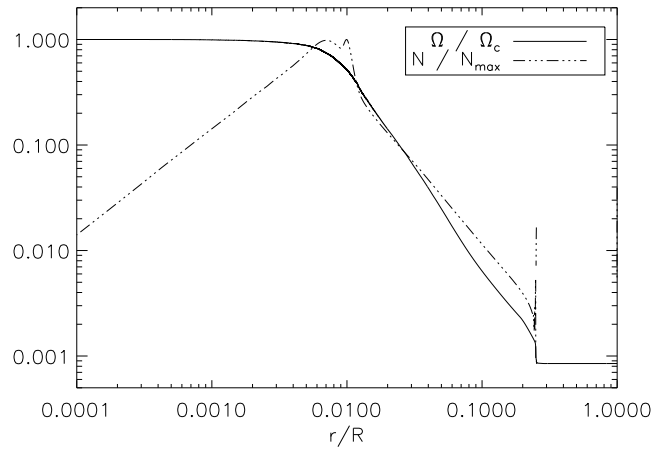}
\caption{Rotation rate normalised by its central value (solid line) and buoyancy frequency normalised by its maximum value (dashed dots lines) versus normalised radius for the model described in Sect.~\ref{models} (logarithmic scale in both axes). 
\label{rotation}}
\end{center}
\end{figure}

The equilibrium model was computed with the stellar evolution code CESTAM  \citep{Marques2013}. The atmosphere was computed assuming a grey Eddington approximation. Convection was included according to \cite{CGM1996}, with a mixing-length parameter $\alpha=0.67$.  The initial chemical composition was the solar composition of \cite{Asplund05}, with a helium mass fraction $Y=0.261$ and metallicity $Z=0.0138$, and diffusion was not included. We used the OPAL equation of state
\citep{Rogers96} and opacities \citep{Iglesias96}, complemented, at $T < 10^4$ K, by the \cite{Alexander94} opacities. We adopted the NACRE nuclear reaction rates \citep{Angulo99} except for the $\element[][14]{N}  + \element{p}$ reaction, where we used the reaction rates of \cite{Imbriani04}. The model was computed without rotation.

To obtain a rotation profile as realistic as possible, we computed an equivalent  model with rotation \citep[using the procedure  described in][]{Marques2013} and rescaled the rotation profile to gain a core rotation rate  $\Omega_c / (2 \pi) = 1 \, \mu$Hz. This value is consistent with the mean central rotation rate of evolved low-mass stars derived from \emph{Kepler} observations \cite[see][]{Mosser2012,Deheuvels2014}. The resulting rotation profile and buoyancy frequency are shown in Fig.~\ref{rotation}. Consistently with the slow rotation limit we consider, this procedure is equivalent to assuming that rotation has a negligible effect on the star equilibrium structure. The equivalent model including rotation was computed using the same input physics and parameters as the model without rotation. It has a radius as close as possible to the radius of the model without rotation and is in the same evolutionary phase. The main characteristics of the models are given in Table~\ref{tab:mods}. 
 
\begin{table}
   \centering
    \caption{Characteristics of the models computed with rotation and without rotation for comparison.}
  \begin{tabular}{lcccc} 
      \hline
      Model    & $R/R_{\odot}$ & $L/L_{\odot}$ & $T_{\rm eff}$~(K) & Age (Myr)\\
      \hline
      Rotating      & 3.558 & 6.830 & 4952 & 4716  \\
      Non-rotating & 3.555  & 6.784  & 4945 & 4590 \\
      \hline
   \end{tabular}
   \label{tab:mods}
\end{table}
 
Finally, we used the ADIPLS code for adiabatic oscillations \citep{JCD08,JCD11} to compute eigenfunctions and eigenfrequencies. The effect of rotation was also neglected in computing the eigenfunctions and eigenfrequencies.

\subsection{Effect of prograde and retrograde mixed modes}
\label{results_one_mode}

We will show in Paper II that the wave fluxes in the mean energy equation, Eq.~\eqref{energy_shellular_vertical}, is negligible. Therefore, within the TEM formalism, the contribution of the mixed modes to the mean flow consists entirely of the wave flux in the mean angular momentum equation (see Eq.~\ref{shellular_omega_r}). Thus, for ease of notation, we define the rate of temporal variation of the mean specific angular momentum induced by mixed modes by
\begin{align}
\label{Jdot}
\dot{J} = -\frac{1}{r^2} \derivp{}{r} \left( r^2 \mathcal{F}_{\rm waves} \right) \, ,
\end{align}
where we recall that $\mathcal{F}_{\rm waves}$ is given by \eq{shellular_omega_r_wave}. Note that $\dot{J}$ is normalised so that for a given mode of angular degree $\ell$ and azimuthal order $m$ one has $\vert \xi_r^{\ell,m}/R \vert=1$ at the photosphere. 

\begin{figure}[!]
\begin{center}
\includegraphics[width=9.2cm]{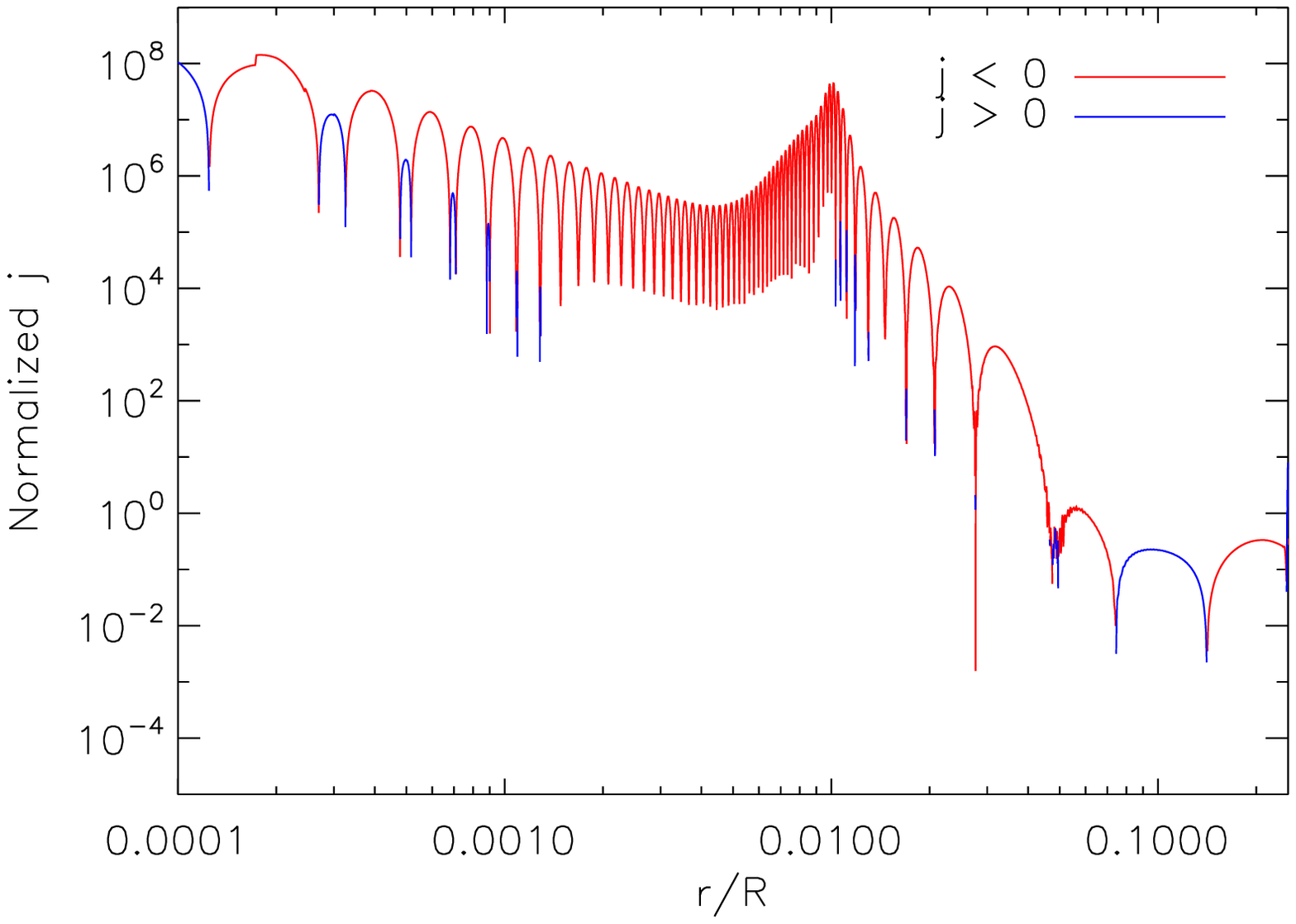}
\includegraphics[width=9.2cm]{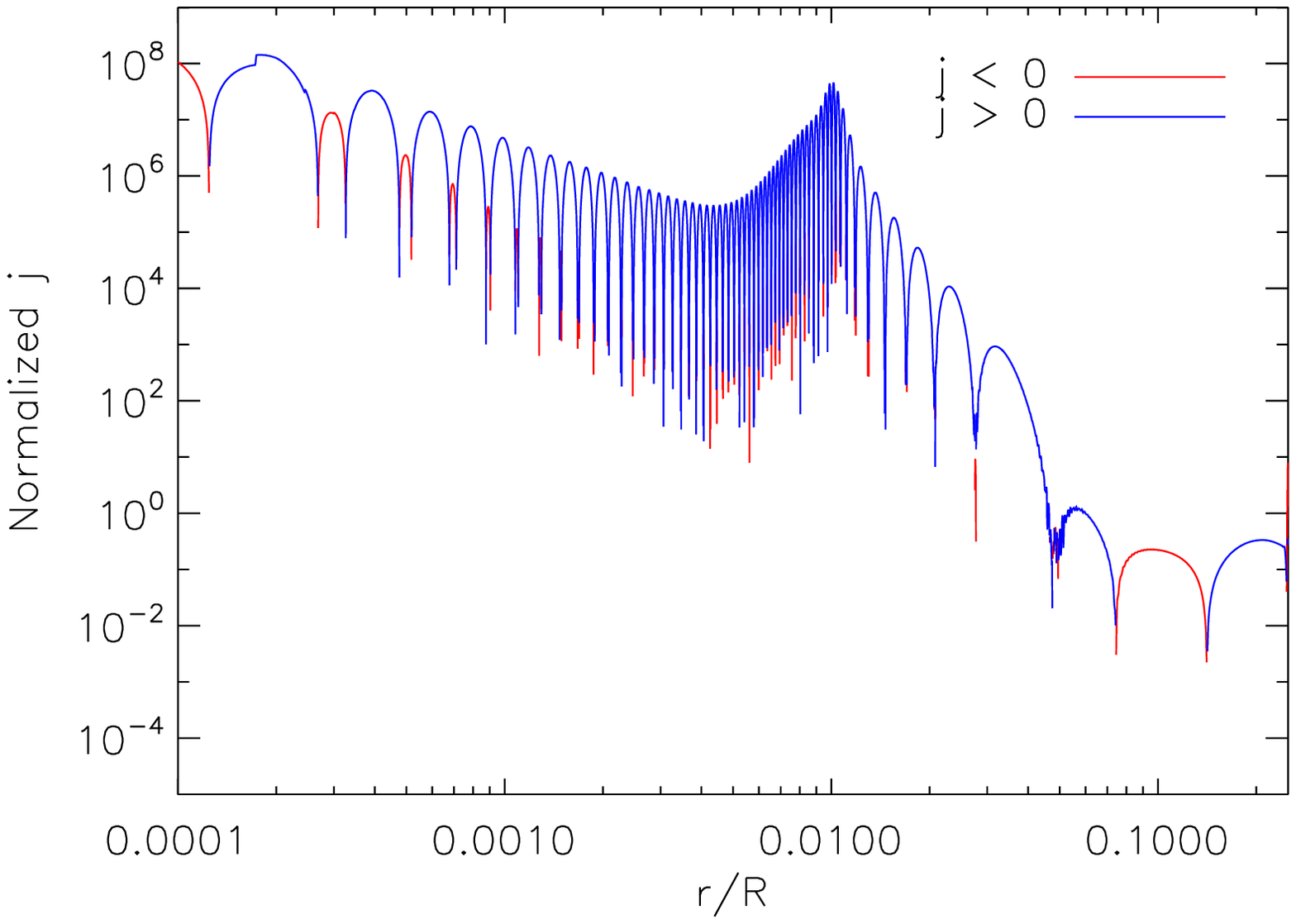}
\includegraphics[width=9.2cm]{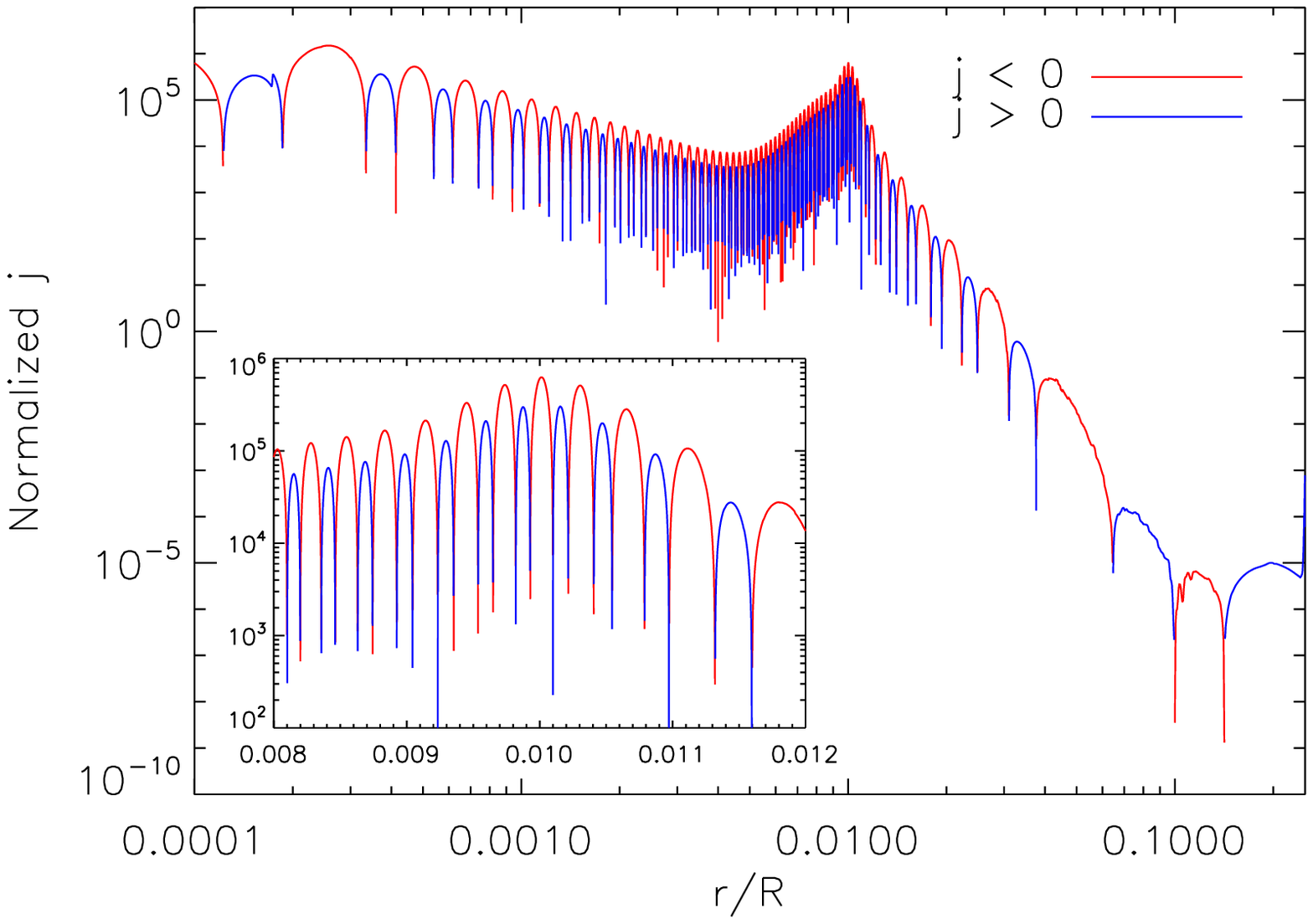}
\caption{Rate of angular momentum transport by mixed modes ($\dot{J}$ as defined by Eq.~\ref{Jdot}) computed as described in Sect.~\ref{sourceterms} for the $1.3 M_\odot$ model described in Sect.~\ref{models} versus the radius normalised by the radius of the radiative region. $\dot{J}$ is normalised so that  $\vert \xi_r^{\ell,m}/R \vert=1$ at the photosphere. 
 The colour code is as follows: the solid red line corresponds to $\dot{J} < 0$ (thus to a spin down) and the solid blue line to  $\dot{J} > 0$ (thus to a spin up). The top panel exhibits the result for the prograde mode $\ell=-m=2$ of frequency $\nu_R = 300,9 \,  \mu$Hz, the middle panel corresponds to the retrograde mode $\ell=m=2$, and the bottom panel to summation of $\dot{J}$ for the prograde and retrograde modes. We only show the inner radiative region, for which our formalism is valid. 
\label{plots_results}}
\end{center}
\end{figure}

Computing $\dot{J}$ as described in Sect.~\ref{sourceterms} leads to the conclusion that prograde modes ($m <0$) extract angular momentum in the regions near the maximum of the buoyancy frequency and the maximum of the rotation rate, therefore slowing down the core. Conversely, the retrograde modes ($m>0$) tend to spin up the core. 
This is illustrated in Fig.~\ref{plots_results} (top and middle panels) for the sectoral modes $\ell=2$, $m=\{-2,+2\}$ at the frequency $\nu_R = \sigma_R / (2\pi) = 300.9 \, \mu$Hz, but this conclusion is valid for the other modes as well. When we consider unstable modes, the opposite situation is found, that is, the retrograde modes extract angular momentum \citep[e.g.,][]{Ando1986,Lee1993,Townsend2008,Townsend2014,Lee2014}. In this case, the wave stresses in the mean angular momentum equation are dominated by transient terms. 

We have also found that, despite an important rotation gradient in the hydrogen burning shell, the dominant contribution to the angular momentum transport comes from the term $\mathcal{D}_\ell^m$ so that, from \eq{final_F_waves} and for a mode of a given angular and azimuthal degree, we can write
\begin{align}
\label{Jdot_approx}
\dot{J} \approx m \hat \sigma \, a_{\ell,m}^2 \, \mathcal{D}_\ell^m \, , 
\end{align}
where $\mathcal{D}_\ell^m$ is given by \eq{coeffDf}.  

To proceed, we considered the sum of the contributions of a prograde and a retrograde mixed modes to the mean angular momentum profile. In the absence of rotation, both prograde and retrograde modes transport angular momentum, but the sum vanishes so that there is no net transport. Including rotation, the net transport of angular momentum is a small residual (see Fig.~\ref{plots_results}  bottom panel). 
The asymmetry between prograde and retrograde modes arises from the Doppler shift terms in the mode frequencies (\emph{i.e.}, $\hat \sigma = \sigma_R + m\Omega_0$). More precisely, the dominant term of \eq{Jdot_approx} comes from the first term of $\mathcal{D}_\ell^m$ in \eq{coeffDf}. This term originates in the second term of \eq{shellular_omega_r_wave}, thus from the wave heat flux. To illustrate it, we added a prograde and retrograde mode, expand in $\Omega_0 / \sigma_R$, and only kept the leading term to obtain
\begin{align}
\dot{J}(\ell,-\vert m \vert) + \dot{J}(\ell,\vert m \vert) \approx 2 \vert m \vert^2 \rho k_r^2 a_{\ell,\vert m \vert}^2  \vert \xi_r^{\ell,\vert m \vert } \vert^2 \left(\frac{\Omega_0}{\sigma_R}\right) \left(\frac{N}{\sigma_R}\right) \, \alpha \, , 
\end{align}
which is negative, since the term $\alpha$ (see Eq.~\ref{alpha_main}) is negative. The sum of the contributions of prograde and retrograde mixed modes thus implies an extraction of angular momentum. 
Nonetheless, in the regions where the contribution of the prograde mode has nodes, the contribution of the retrograde mode dominates due to the second and third terms of $\mathcal{D}_\ell^m$ in \eq{coeffDf}. 
It therefore results in a complex situation where there is a spatial alternation between regions where mixed modes decrease and increase the mean momentum, as shown in Fig.~\ref{plots_results} (bottom panel).  However, before reaching a conclusion, we have to consider the collective effect of mixed modes. 

\subsection{Collective influence of mixed mode}
\label{results_collectivemodes}

\begin{figure}[t]
\begin{center}
\includegraphics[width=9.2cm]{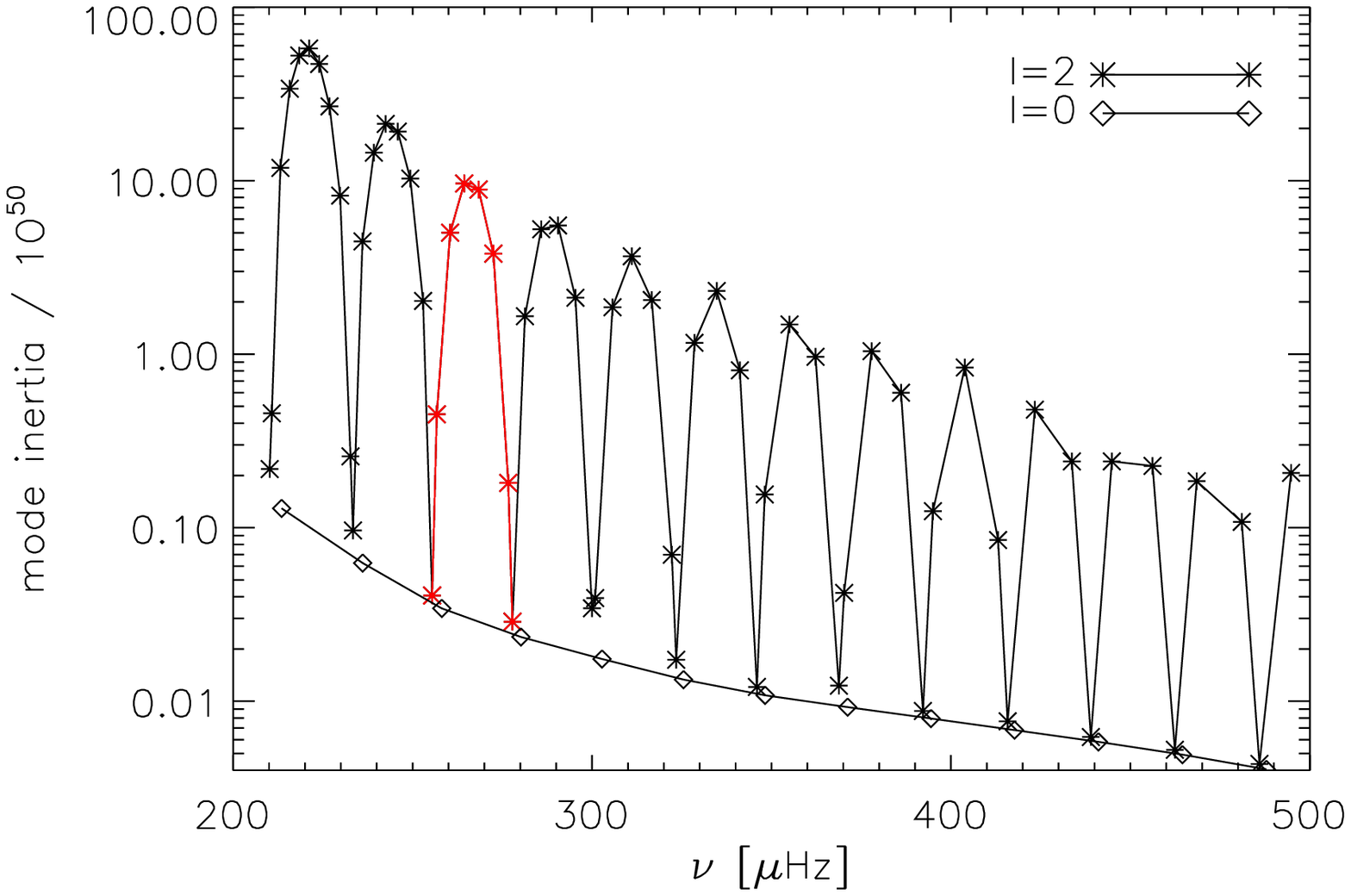}
\includegraphics[width=9.2cm]{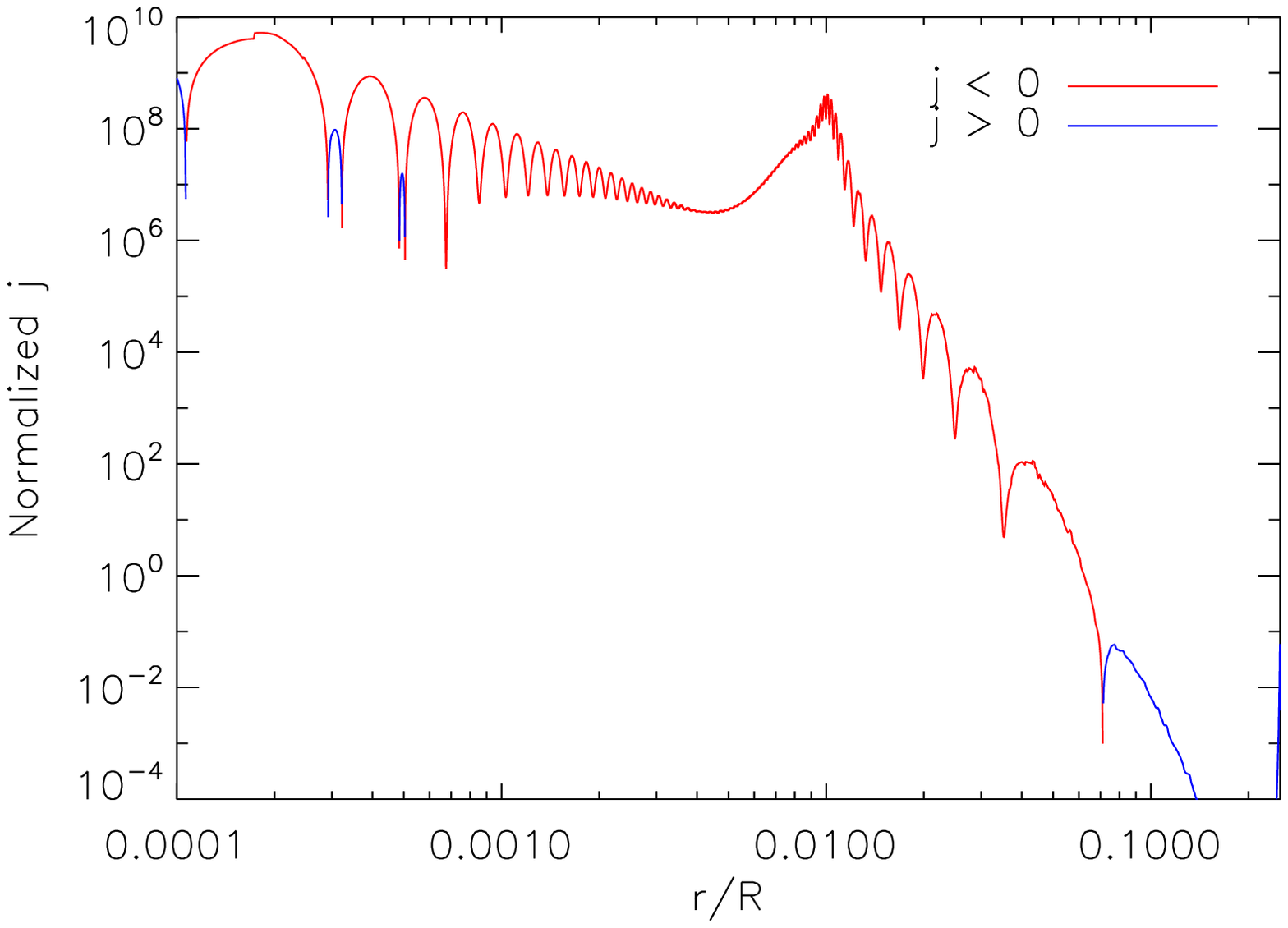}
\caption{{\bf Top panel:} Mode inertia versus mode frequencies for angular degrees $\ell=0$ and $\ell=2$. Mode inertias are computed by integrating the squared eigendisplacement over the stellar mass. {\bf Bottom panel:}  Same as Fig.~\ref{plots_results}, except that $\dot{J}$ is computed by considering the total contribution of the $\ell=2$ modes for the selected range displayed in red in the top panel.
 \label{plots_results_sum}}
\end{center}
\end{figure}

Oscillation spectra of low-mass evolved stars exhibit quasi-periodic patterns with periods that correspond to the large separation as depicted in Fig.~\ref{plots_results_sum} (top panel). Each pattern is made of p-dominated mixed modes separated by several g-dominated modes. It is therefore instructive to consider the collective effect of mixed modes on the mean angular momentum. 

We  selected a range of modes, shown in red in Fig.~\ref{plots_results_sum} (top panel), and computed the total contribution to $\dot{J}$ by summing the contribution of all modes belonging to this range, including the different values of the azimuthal order $m$  for each mode. Given the spatial shift between the consecutive mixed modes, the summation induces a smoothing of $\dot{J}$, as shown by Fig.~\ref{plots_results_sum} (bottom panel). The collective effect of mixed modes is  to decrease the mean angular momentum and thus to slow down the core rotation of the star. 

Nevertheless, we emphasise that to realistically compute the effect of the modes on the mean angular momentum profile, we need to  consider all modes with realistic individual amplitudes This is the scope of paper II. 

\section{Conclusion}
\label{conclusions}

  Our aim here was to establish a formalism allowing us to calculate the transport of angular momentum by mixed modes  in the inner radiative region of evolved low-mass stars. We thus considered the transformed Eulerian mean (TEM) formalism that allowed us to take into account the combined effect of both the wave momentum flux in the mean angular momentum equation and the wave heat flux in the mean energy equation. Indeed, the effect of the wave heat flux on the mean flow is generally neglected in the context of angular momentum transport by waves in stars. Here, the TEM allows us to include its impact on the mean angular momentum in a relatively simple way. 
  
  Given the important contraction of the core and expansion of the envelope of low-mass stars after the main sequence, a strong radial rotation gradient appears. This was taken into account by adopting the shelullar approximation. While a two-dimensional treatment is certainly desirable and more accurate \citep[e.g.,][]{Espinosa2013}, this approximation provides a first approach of the problem while keeping it tractable.  
   
 Subsequently, the wave field has been modelled so as to account for the transport of angular momentum by mixed modes. We limited our investigation to the inner radiative regions. Thus, several simplifications are justified;  the low-frequency limit (\emph{i.e.} $\sigma^2_R \ll \{N^2, S_\ell^2\}$), the quasi-adiabatic approach (the local thermal time-scale is higher than the modal period), and the low-rotation limit as suggested by recent asteroseismic observations \citep{Deheuvels2012,Mosser2012}. It allowed us to obtain an explicit expression of the wave-related terms appearing in the mean angular momentum equation. 
 
Then, the formalism was applied to a benchmark model of a $1.3 M_\odot$ evolved star.  We found that the influence of the wave heat flux on the mean angular momentum is not negligible and even dominant when considering the sum of prograde and retrograde modes. In addition, we showed that the overall effect of mixed modes is to extract angular momentum in the innermost radiative layers.  

 Nevertheless, more realistic estimates of mode amplitudes are required for a more quantitative determination of the amount of angular momentum transported by mixed modes. The  second paper of this series is dedicated to this. 
 
\begin{acknowledgements}
We acknowledge the ANR (Agence Nationale de la Recherche, France) program IDEE (n$^\circ$ANR-12-BS05-0008) ``Interaction Des \'Etoiles et des Exoplan\`etes'' as well as financial support from "Programme National de Physique Stellaire" (PNPS) of CNRS/INSU, France. RMO aknowledges funding for the Stellar Astrophysics Centre, provided by The Danish National Research Foundation, and funding for the ASTERISK project (ASTERoseismic Investigations with SONG and Kepler) provided by the European Research Council (Grant agreement no.: 267864).
\end{acknowledgements}

\bibliographystyle{aa}
\bibliography{bib.bib}


\appendix 
\section{Derivation of the wave flux in the mean angular momentum equation (Eq.~\ref{shellular_omega_r})}
\label{AppA}

We assume that the perturbations consist in non-radial oscillations and that they behave linearly as $\exp\left[ i \left( \sigma t + m \phi \right) \right]$, where $\sigma$ is the modal frequency, $t$ the time, and $m$ the azimuthal degree.  Consequently, for a given quantity $f$, its perturbation can be written  
\begin{align}
\label{def_perturbation}
f^\prime = a \, \mathcal{R}_e \left\{ \tilde{f}^\prime(r,\theta) \; e^{i (\sigma_R t + m \phi)}\right\}\, ,
\end{align} 
where $a$ is the amplitude, $\sigma_R$ the real part of the frequency $\sigma$, and $\mathcal{R}_e$ stands for the real value. From \eq{def_perturbation} one immediately deduces a relation for squared quantities that reads
\begin{align}
\label{squared_perturbation}
2 \overline{f^\prime g^\prime} = a^2 \, \mathcal{R}e \left\{\tilde{f}^\prime \tilde{g}^{\prime \ast} \right\} \, .
\end{align}

One can then express the Eulerian perturbation of the entropy in terms of the Lagrangian one, using $\delta s = s^\prime + \xi_r \; {\rm d}\left<s\right>/{\rm d}r$, where $\xi_r$ is the radial component of the wave displacement. It allows us to rewrite the divergence of the wave flux in the momentum equation (see Eq.~\ref{energy_shellular_vertical}) such as 
\begin{align}
\label{source_terms_app}
&-\frac{1}{r^2} \derivp{}{r} \left( r^2 \mathcal{F}_{\rm waves}\right) =  - \frac{1}{2 r^2} \derivp{}{r}\mathcal{R}_e \left\{ \left< a^2 r^2 \varpi \,\left< \rho \right> \, \mathcal{S}_1  \right>  \right\}  
\end{align}
with
\begin{align}
\label{S1_app}
\mathcal{S}_1 = \tilde{\vt}_r^\prime \tilde{\vt}_\phi^{\prime \ast} 
- 2 \cos \theta \, \Omega_0 \left[ \tilde{\vt}_\theta^{\prime} \tilde{\xi}_r^{\prime \ast} - \tilde{\vt}_\theta^{\prime} \delta s^{\ast} \left(\deriv{\left< s\right>}{r}\right)^{-1} \right] \, .
\end{align}
In the following, our objective is to provide a general expression of \eq{source_terms_app} using the full set of equations governing waves. Note that in the following, the tilde will be omitted for the perturbation  as well as overbar and brackets for the unperturbed quantities. 

\subsection{Using the full wave equations for expressing \eq{S1_app}}

In this subsection, we aim to express the correlation products appearing in \eq{S1_app} using the set of \eq{eq1_stationaire} to \eq{eq4_stationaire}. 

\subsubsection{Derivation of the first term of \eq{S1_app}}
\label{first_S1}

First, Eqs.~(\ref{eq2_stationaire}) and (\ref{eq3_stationaire}) are inserted into (\ref{eq4_stationaire}), and an integration by part is performed. Second, Eqs.~(\ref{eq1_stationaire}) to   (\ref{eq3_stationaire}) are used as well as the relation
\begin{align}
\label{intermediate3}
 & \frac{\vt_\phi^{\prime \ast} \rho \vec \vt_\bot^{\prime}}{\varpi \vert \hat \sigma \vert^2} \hat \sigma  \nabla_\bot h 
 = \nonumber \\
 &\frac{2 m}{\hat \sigma^\ast} \frac{h}{\varpi^2} \rho \vt_\phi^{\ast} \left( \vec \vt_\bot^\prime \cdot \nabla_\bot \varpi \right) 
 - \varpi \rho \vt_\phi^{\prime \ast} \vec \vt_\bot^\prime \cdot \nabla_\bot \left( \frac{1}{\hat \sigma^\ast} \right) \, .
\end{align}
Finally, taking the real part and considering that the imaginary part of the frequency vanishes ($\sigma_I = 0$) since we consider steady waves (see Sect.~\ref{sourceterms}), one gets   
\begin{align}
\label{moment_app}
&- \nabla_\bot \cdot \left[ \rho \varpi\overline{ \vec \vt_\bot^\prime \vt_\phi^\prime} \right]  = 
- \frac{m\rho}{ 2} \left[ \frac{\left(\vec \vt_\bot^\prime \cdot \rho^{-1} \vec \nabla_\bot p \right) \vec \vt_\bot^{\prime \ast} \cdot \vec A}{\hat \sigma^2} \right]_I  \nonumber \\
&- \frac{m\rho}{2 \varpi} \left( \frac{\rho  \varpi^2}{m^2 \Gamma_1 p} - \frac{1}{\hat \sigma^2} \right) \left[ \hat \sigma \vec \vt_\bot^\prime \vt_\phi^{\prime \ast} \right] _R\cdot \vec \nabla_\bot h \nonumber \\
&- \frac{\rho}{2} \left[ \frac{\left( \vec \vt_\bot^\prime \cdot \vec A + \vec \vt_\bot^\prime \cdot \vec \nabla_\bot  \right) \left(\vec \vt_\bot^{\prime \ast} \cdot \vec \nabla_\bot h \right)}{\hat \sigma }\right]_I 
+ \frac{\rho \varpi}{2}  \left[ \vt_\phi^{\prime \ast} \hat \sigma \rho_T \frac{\delta s}{c_p} \right]_I\nonumber \\
&+ \frac{m \rho}{2} \left[ \frac{\left(\vec \vt_\bot^{\prime} \cdot \rho^{-1} \vec \nabla_\bot p \right) \rho_T \delta s^{\ast} / c_p }{\hat \sigma}\right]_R \, , 
\end{align}
where we introduced the notation $\hat \sigma = \sigma_R + m \Omega_0$, $\rho_T=-\left(\partial \ln \rho / \partial \ln T \right)_p$, and the subscripts $I$ and $R$ denote the imaginary and real part, respectively. 
Note that this expression corresponds to Eq. (36.16) of \cite{Unno89} and was first derived by \cite{Ando1983}. 

To go further, we assume shellular rotation and that the equilibrium state is not deformed by rotation. Moreover, integration over the solid angle is performed so that it finally gives 
\begin{align}
\label{first_S1_eq}
&- \frac{1}{2 r^2} \derivp{}{r} \left< r^2 \varpi \rho \vt_r^\prime \vt_\phi^{\prime \ast} \right>_R  
=  
- \frac{\rho r}{2 m c_s^2} \derivp{\left(r^2 \Omega_0 \right)}{r} \hat \sigma \left< \sin^3 \theta  \; \vt_r^\prime \vt_\phi^{\prime \ast}  \right>_R \nonumber \\
&  - \frac{\rho r^2 \Omega_0}{m c_s^2}  \hat \sigma \left< \sin^2 \theta \cos \theta \; \vt_\theta^\prime \vt_\phi^{\prime \ast}  \right>_R 
+ \frac{m \rho \Omega_0}{\hat \sigma} \left< \cos \theta \; \vt_\theta^\prime \vt_\phi^{\prime \ast}  \right>_R \nonumber \\
&+ \frac{m \rho}{2 r \hat \sigma}  \derivp{\left(r^2 \Omega_0 \right)}{r} \left< \sin \theta  \; \vt_r^\prime \vt_\phi^{\prime \ast}  \right>_R 
-\frac{\rho (Ar-1) \Omega_0}{\hat \sigma}  
\left< \cos \theta \sin \theta \; \vt_r^\prime \vt_\theta^{\prime \ast} \right>_I  \nonumber \\
&-\frac{\rho r \Omega_0}{\hat \sigma}
\left<  \cos \theta \sin \theta \, \vt_r^\prime \derivp{\vt_\theta^{\prime \ast}}{r}\right>_I 
-\frac{\rho}{2 r \hat \sigma} \derivp{\left(r^2 \Omega_0\right)}{r}
\left< \sin^2 \theta \, \vt_\theta^\prime \derivp{\vt_r^{\prime \ast}}{\theta}\right>_I \nonumber \\
&+ \frac{\rho r \rho_T}{2} \hat \sigma\left< \sin \theta \, \frac{\delta s}{c_p} \vt_\phi^{\prime \ast} \right>_I 
+ \frac{m \rho_T}{2 \hat \sigma} \derivp{p}{r} \left< \vt_r^\prime \frac{\delta s^{\ast}}{c_p} \right>_R \, .
\end{align}

\subsubsection{Derivation of the second term of \eq{S1_app}}
\label{second_S1}

We start from \eq{eq2_stationaire}, which immediately gives 
\begin{align}
\label{C2}
2\Omega_0 \cos \theta \, \vt_\theta^\prime &= -\frac{imp^\prime}{\varpi \rho} - i \hat \sigma \vt_\phi^\prime 
- \left(\frac{\sin\theta}{r}\right) \derivp{\left(r^2\Omega_0\right)}{r} \vt_r^\prime \, .
\end{align}
Equation \eq{C2} is further multiplied by $r^2 \rho \varpi \xi_r^{\ast}$ and radial derivation is performed to give
\begin{align}
\label{C3}
 & \frac{1}{r^2} \derivp{}{r} \left( 2 \Omega_0 r^2 \varpi \rho \cos \theta  \vt_\theta^{\prime} \xi_r^{\prime \ast}  \right)  = 
- i m \left[ \frac{p^\prime}{r^2} \derivp{\left(r^2 \xi_r^{\ast}\right)}{r} + \xi_r^{\ast} \derivp{p^\prime}{r}\right] \nonumber \\ 
&-  i \varpi \rho \hat \sigma \left( \vt_\phi^\prime \derivp{\xi_r^{\ast}}{r} + \xi_r^{\ast} \derivp{\vt_\phi^\prime}{r}\right) - \rho \sin^2\theta \derivp{^2\left(r^2 \Omega_0\right)}{r^2}  \vt_r^\prime \xi_r^{\ast}
\nonumber \\
&- \rho \sin^2\theta \derivp{\left(r^2 \Omega_0\right)}{r} \left[ \left( \frac{2}{r} + \derivp{\ln \rho}{r}  \right) \vt_r^\prime \xi_r^{\ast} + \left( \vt_r^\prime \derivp{\xi_r^{\ast}}{r} + \xi_r^{\ast} \derivp{\vt_r^\prime}{r} \right)\right]
 \nonumber \\
&- i \sin \theta \rho \hat \sigma  \vt_\phi^\prime \xi_r^{\ast} 
\left[ 1 + \derivp{\ln \rho}{\ln r} +  \frac{2\sigma_R}{\hat \sigma} +  \frac{m}{r \hat \sigma} \derivp{ \left(r^2\Omega_0\right)}{ r} \right] \, .
\end{align}
 Now one needs to express both the perturbation of pressure and its radial derivative. To this end, we use \eq{eq3_stationaire} and \eq{eq4_stationaire} as well as Eqs.~(\ref{eq1_stationaire}) to (\ref{eq3_stationaire}) to get 
\begin{align}
\label{p_prime}
&\frac{p^\prime}{\Gamma_1 p} = -\frac{i \vt_r^\prime A}{\hat \sigma}
+ \frac{i  \nabla \cdot \left(\rho \vec \vt_\bot^\prime \right)}{\rho \hat \sigma}
- \frac{m \vt_\phi^\prime}{\varpi \hat \sigma} 
+ \rho_T \frac{\delta s}{c_p}
\end{align}
and
\begin{align}
\label{gradp_prime}
&\derivp{p^\prime}{r} = - i \left[ \rho \hat \sigma - \frac{\sin^2\theta}{m c_s^2} \derivp{p}{r}  \derivp{\left( r^2 \Omega_0 \right)}{r} - \frac{A}{\hat \sigma} \derivp{p}{r} \right] \vt_r^\prime  \\
&+ \left[2 \rho \Omega_0 \sin \theta - \frac{\varpi \hat \sigma}{m c_s^2} \derivp{p}{r}  \right] \vt_\phi^\prime 
+ 2 i \derivp{p}{r}  \frac{r \sin \theta \cos \theta \Omega_0}{m c_s^2} \vt_\theta^\prime 
 \nonumber 
- \rho_T \derivp{p}{r} \frac{\delta s}{c_p} \, .
\end{align}

Finally, inserting Eqs.~(\ref{gradp_prime}) and (\ref{p_prime}) into \eq{C3}, integrating over the solid angle and taking the real part gives 
\begin{align}
\label{second_S1_eq}
 &\frac{1}{2 r^2} \derivp{}{r} \mathcal{R}_e \left< 2 \Omega_0 r^2 \varpi \rho \cos \theta  \vt_\theta^{\prime} \xi_r^{\prime \ast}  \right>_R  = 
\nonumber \\
& \frac{\rho}{2} \Bigg[  A r + 3  
 + \frac{m}{r \hat \sigma} \derivp{ \left(r^2\Omega_0\right)}{ r} \Bigg] \left<\sin \theta\; \vt_\phi \vt_r^{\ast}\right>_R 
 - \frac{m \rho_T}{2 \hat \sigma}\derivp{p}{r} \left< \frac{\delta s}{c_p} \vt_r^{\ast}\right>_R  
 \nonumber \\
 &+ \frac{m \rho c_s^2}{2 r^3 \hat \sigma}  \left<\frac{1}{\sin \theta} \derivp{\left(\sin \theta \vt_\theta \right)}{\theta}\;  \derivp{\left(r^2 \xi_r^{\ast}\right)}{r}\right>_R  + \frac{m}{2 r^2} \rho_T \Gamma_1 p \, \left< \frac{\delta s}{c_p} \derivp{\left( r^2 \xi_r^{\ast}\right)}{r}\right>_I
 \nonumber \\
&+ \frac{\rho r}{2} \hat \sigma \left< \sin \theta \left[ \vt_\phi^\prime \derivp{\xi_r^{\ast}}{r} + \derivp{\vt_\phi^\prime}{r} \xi_r^{\ast}\right] \right>_I 
- \frac{m^2 \rho c_s^2}{2 r^3 \hat \sigma}  \left< \frac{1}{\sin \theta} \vt_\phi^\prime \derivp{\left(r^2 \xi_r^{\ast}\right)}{r} \right>_I  \nonumber \\
 &- \frac{ \rho \Omega_0}{\hat \sigma} \frac{1}{\Gamma_1} \derivp{\ln p}{\ln r} 
 \left< \cos \theta \sin \theta \; \vt_\theta^\prime \vt_r^{\ast} \right>_I \, .
\end{align}

\subsubsection{Derivation of the third term of \eq{S1_app}} 
\label{third_S1}

The derivation is similar to the previous subsection, it gives 
\begin{align}
\label{third_S1_eq}
 &- \frac{1}{2 r^2} \derivp{}{r} \left< 2 \Omega_0 r^2 \varpi \rho \cos \theta  \vt_\theta^{\prime} \xi_s^{\prime \ast}  \right>_R = \frac{m \rho}{2} \hat \sigma \left[ 1-\frac{A}{\rho \hat \sigma^2 }\derivp{p}{r}\right] \left< \vt_r^\prime   \xi_s^{\ast} \right>_R \nonumber \\
& -\frac{\rho}{2} \hat \sigma \left[ Ar+3+\frac{m}{r \hat \sigma} \derivp{\left(r^2 \Omega_0\right)}{r}\right]  \left<  \sin \theta \, \vt_\phi^\prime \xi_s^{\ast} \right>_I \nonumber \\
&+ \frac{\rho}{2} \left[(Ar+2) \frac{1}{r} \derivp{\left(r^2 \Omega_0\right)}{r} + \derivp{^2 \left(r^2 \Omega_0\right)}{r^2} \right] \left< \sin^2\theta \, \vt_r^\prime   \xi_s^{\ast} \right>_R \nonumber \\
&+ \frac{\rho}{2}  \derivp{\left(r^2 \Omega_0\right)}{r}  \left< \sin^2 \theta \, \left( \vt_r^\prime \derivp{\xi_s^{\ast}}{r} + \xi_s^{\ast}  \derivp{\vt_r^\prime}{r} \right) \right>_R \nonumber \\ 
&- \frac{\rho r}{2} \hat \sigma \,  \left< \sin \theta \, \left( \vt_\phi^\prime \derivp{\xi_s^{\ast}}{r} + \xi_s^{\ast}  \derivp{\vt_\phi^\prime}{r} \right) \right>_I 
+ \frac{m A \Gamma_1 p}{2r^2 \hat \sigma}\left< \vt_r^\prime  \derivp{\left(r^2 \xi_s^{\ast}\right)}{r} \right>_R \nonumber \\
&- \frac{m c_s^2}{2 r^2 \hat \sigma}  \left< \vec \nabla_\bot \cdot \left(\rho \vec \vt_\bot\right)  \derivp{\left(r^2 \xi_s^{\ast}\right)}{r} \right>_R 
+ \frac{m^2 \rho c_s^2}{2 r^3 \hat \sigma}  \left< \frac{1}{\sin \theta} \; \vt_\phi  \derivp{\left(r^2 \xi_s^{\ast}\right)}{r} \right>_I  \nonumber \\
&- \frac{r \Omega_0}{c_s^2} \derivp{p}{r} \,  \left< \sin \theta \cos \theta \, \vt_\theta^\prime \xi_s^{\ast} \right>_R  \, ,
\end{align}
where we have introduced the notations  
\begin{align}
\xi_s = \left(\deriv{s}{r}\right)^{-1} \delta s \, \quad {\rm and} \quad \vt_s^\prime = i \hat \sigma \xi_s \, .
\end{align}

\subsection{The case of low rotation and frequency limit}

As described in the main text (see Sect.~\ref{sourceterms}) we restrict ourselves in the limit of low rotation (\emph{i.e.} $\sigma_R \gg \Omega_0$) and low frequency (\emph{i.e.} $\sigma_R \ll N$, where $N$ is the buoyancy frequency). Those approximations will permit us to derive tractable expressions for Eqs.~(\ref{first_S1_eq}), (\ref{second_S1_eq}), and (\ref{third_S1_eq}). 
To this end, the first step consists in considering the wave equations without rotation and introducing them into Eqs.~(\ref{first_S1_eq}), (\ref{second_S1_eq}), and (\ref{third_S1_eq}). This is equivalent in considering a first-order development in term of rotation for \eq{S1_app}. 

Projection onto the spherical harmonics is thus performed in the limit of low rotation. For one normal mode of a given $m$ and $\ell$, the eigen-displacement and velocity can be decomposed such as 
\begin{align}
\label{decomp}
\vec \xi^m = \xi_r^{\ell,m} \, \mathcal{\vec R}_{\ell,m} + \xi_h^{\ell,m} \,  \mathcal{\vec S}_{\ell,m} \, ,
\end{align}
where \citep{Rieutord1987}
\begin{align}
\mathcal{\vec R}_{\ell,m} =  Y_\ell^m \vec e_r \,, \quad {\rm and} \quad
\mathcal{\vec S}_{\ell,m} =  \left( \derivp{Y_\ell^m}{\theta} \vec e_\theta + \frac{1}{\sin \theta} \derivp{Y_\ell^m}{\phi} \vec e_\phi \right) \;, \\ 
\end{align}
so that 
\begin{align}
\xi_r^\prime =  \, \xi_r^{\ell,m} \, Y_\ell^m \,,\;  
\xi_\theta^\prime = \,  \xi_h^{\ell,m} \, \derivp{Y_\ell^m}{\theta} \,,\; 
\xi_\phi^\prime =  \,  \xi_h^{\ell,m} \, \frac{1}{\sin \theta} \derivp{Y_\ell^m}{\phi}  \, ,
\end{align}
and 
\begin{align}
\vt_r^\prime = i \, u_\ell^m \, Y_\ell^m \,,\;  
\vt_\theta^\prime = i \, \vt_\ell^m \, \derivp{Y_\ell^m}{\theta} \,, \; 
\vt_\phi^\prime =   i \, \vt_\ell^m \, \frac{1}{\sin \theta} \derivp{Y_\ell^m}{\phi}  \, .
\end{align}
where we used $\vec v^m = i \hat \sigma \vec \xi^m$. 

In the  non-rotating limit and providing the decomposition given by \eq{decomp}, using Eqs.~(\ref{eq1_stationaire}) to (\ref{eq4_stationaire}), $\vt_\ell^m$ can be expressed as a function of $\delta s_{\ell}^m$ and $u_\ell^m$ only through the relation 
\begin{align}
\label{vellm_norot}
\vt_\ell^m = r \, \mathcal{G} \, \left[\sigma_R \rho_T \frac{\delta s_{\ell}^m}{c_p} -\frac{1}{r^2} \deriv{\left( r^2 u_\ell^m\right)}{r} -  \frac{u_\ell^m}{\Gamma_1} \deriv{\ln p}{r}\right]
\end{align}
with  
\begin{align}
\mathcal{G} = \frac{c_s^2}{r^2 \, \sigma_R^2} \left[1- \frac{S_\ell^2}{\sigma_R^2} \right]^{-1} \, , {\rm and} \quad
S_\ell^2 = \ell(\ell+1) \, \frac{c_s^2}{r^2} \, .
\end{align}

To go further, in the asymptotic, quasi-adiabatic limit ($\sigma_R \ll N$), $\delta s_{\ell}^m$ is a function of $u_\ell^m$ (or equivalently $\xi_r^{\ell,m}$) so that $\vt_\ell^m$ depends only on $u_\ell^m$. It reads \citep[see][for details]{Godart2009}
\begin{align}
\label{deltaS_final}
i \sigma_R \delta s_\ell^m &= \frac{L}{4 \pi r^2 \rho T} \left(\frac{\nabla_{\rm ad}}{\nabla}-1\right) k_r^2 \xi_r^{\ell,m}  
\end{align}
and
\begin{align}
\label{incompressibility}
\deriv{\xi_r^{\ell,m}}{r} \simeq \frac{\ell (\ell+1)}{r} \xi_h^{\ell,m} \, ,
\end{align}
where $L$ is the luminosity, $T$ the temperature, $\rho$ the density, $\nabla$ the temperature gradient, and $\nabla_{\rm ad}$ the adiabatic temperature gradient. 

Finally, after projection onto the spherical harmonics and using \eq{vellm_norot}, \eq{deltaS_final}, and \eq{incompressibility}, one obtains for a mode of a given angular degree ($\ell$) and azimuthal degree ($m$) an expression of the form 
\begin{align}
\label{final_F_waves_app}
&-\frac{1}{r^2} \derivp{}{r} r^2 \mathcal{F}_{\rm waves} = 
\\
&a_{\ell,m}^{2} \left[ \mathcal{A}_{\ell}^m \; \derivp{^2\left(r^2 \Omega_0 \right)}{r^2} +\mathcal{B}_{\ell}^m \; \derivp{\left(r^2 \Omega_0 \right)}{r} + \mathcal{C}_{\ell}^m \; r^2 \Omega_0 + m \hat \sigma \, \mathcal{D}_{\ell}^m \right] \nonumber \, , 
\end{align}
with 
\begin{align}
\label{coeffAf}
4 \pi \, \mathcal{A}_{\ell,m} = \frac{\rho \alpha k_r^2}{2} \, K_1 \, \vert \xi_r^{\ell,m} \vert^2 \, ,
\end{align}

\begin{align}
\label{coeffBf}
&\frac{4 \pi \mathcal{B}_{\ell,m}}{\rho k_r^2 \vert \xi_r^{\ell,m} \vert^2} = 
 \frac{1}{2} \alpha \, \left(A+2 \zeta_1+\deriv{}{r}\ln\left[r^2 \zeta_2 \right]\right) \, K_1  \\
&+\frac{ r^2  \sigma_R \hat \sigma}{2 c_s^2} \mathcal{G} \zeta_0 K_1 
+ \frac{m^2}{2} \alpha \mathcal{G} \zeta_3 
- \frac{\sigma_R}{\hat \sigma} \mathcal{G}  \zeta_0 \left( m^2 + \frac{K_0}{2} \right) \nonumber
\end{align}

\begin{align}
\label{coeffCf}
&\frac{4 \pi \, \mathcal{C}_{\ell,m}}{\rho k_r^2 \vert \xi_r^{\ell,m} \vert^2} =  - \frac{\sigma_R}{\hat \sigma}\frac{1}{\Gamma_1 p}\deriv{p}{r} \mathcal{G} K_2 \left( \zeta_0 -  \alpha \zeta_3 \right)  \\ 
&+\frac{\sigma_R}{\hat \sigma} \mathcal{G}  \zeta_0 K_2 \, \left[ \frac{\left(Ar-1\right)}{r} + \deriv{}{r}\ln \left(r \mathcal{G}  k_r^2 \zeta_0\right) + \zeta_1 \right] \, , \nonumber
\end{align}
and
\begin{align}
\label{coeffDf}
&\frac{4 \pi \, \mathcal{D}_{\ell,m}}{\rho k_r^2 \vert \xi_r^{\ell,m} \vert^2} = \frac{\alpha}{2} \left[ 1 - \frac{A}{\rho \hat \sigma^2} \deriv{p}{r} \right]   \\
&- \frac{r^2}{2} \mathcal{G} \zeta_0 \left[ A+\frac{1}{r}+\zeta_1+\deriv{}{r}\ln\left(r\mathcal{G} k_r^2 \zeta_0\right) -\frac{1}{\Gamma_1} \deriv{\ln p}{r}  \right] \nonumber \\
&+ \frac{r^2}{2} \mathcal{G} \alpha \zeta_3  \left[ A +\frac{3}{r}+2 \zeta_1+\deriv{}{r}\ln\left(r\mathcal{G} \sigma_R \zeta_3 \right) + \deriv{}{r}\ln\left(\zeta_2\right) \right] \, ,\nonumber
\end{align}
where for \eq{coeffDf} we use the approximation $\mathcal{G}^{-1}=- \ell(\ell+1)$ obtained in the limit $S_\ell^2 \gg \sigma_R^2$, and 
\begin{align}
\alpha &= -\frac{L}{4 \pi r^2 \rho T} \left( \frac{\nabla_{\rm ad}}{\nabla}-1 \right) \, \left(\deriv{s}{r}\right)^{-1} \nonumber \\
\zeta_0 &= \alpha \frac{\rho_T}{c_p} \left(\deriv{s}{r}\right)\; ; \quad
\zeta_1 = \frac{\ell(\ell+1)}{r} \frac{\xi_h^{\ell,m}}{\xi_r^{\ell,m}} \nonumber \\
\zeta_2 &=  \frac{\alpha k_r^2}{\sigma_R}\; ; \quad
\zeta_3 =  \frac{2}{r} + \zeta_1 + \frac{1}{\Gamma_1} \deriv{\ln p}{r} 
\end{align}
as well as
\begin{align}
K_0 &= \ell^2 \left(J_{\ell+1}^{m}\right)^2 + (\ell+1)^2 \left(J_{\ell}^{m}\right)^2 \nonumber \\
K_1 &= 1-\left(J_{\ell+1}^{m}\right)^2 - \left(J_{\ell}^{m}\right)^2 \nonumber \\
K_2 &= \ell \left(J_{\ell+1}^{m}\right)^2 - (\ell+1) \left(J_{\ell}^{m}\right)^2 \, .
\end{align}

Finally, we note that the normalization condition for spherical harmonics is 
$\left< Y_\ell^m Y_{\ell^\prime}^{m^\prime} \right> = \delta_{\ell,\ell^\prime} \, \delta_{m,m^\prime}$, where $\delta$ is the Kronecker symbol. Note also that the following relation have been used
\begin{align}
&\cos \theta \; Y_\ell^m = J_{\ell+1}^m Y_{\ell+1}^m + J_{\ell}^m Y_{\ell-1}^m \, , \\
&\sin \theta \; \derivp{Y_\ell^m}{\theta} = \ell J_{\ell+1}^m Y_{\ell+1}^m - (\ell+1) J_{\ell}^m Y_{\ell-1}^m \, , \\
&\derivp{^2 Y_\ell^m}{\theta^2} + \frac{\cos \theta}{\sin \theta} \derivp{Y_\ell^m}{\theta} + \frac{1}{\sin^2 \theta}  \derivp{^2 Y_\ell^m}{\phi^2} = -\ell(\ell+1) Y_\ell^m \, , 
\end{align}
where 
\begin{align}
\label{defs_coeffs}
J_\ell^m &= \left[\frac{\ell^2-m^2}{(4\ell^2-1)}\right]^{1/2} \, ,  
\end{align}
if $\ell > \vert m \vert$, and $J_\ell^m=0$ otherwise. 

\end{document}